\documentclass[aps,prl,twocolumn,superscriptaddress]{revtex4-2}
\usepackage{graphicx}  % needed for figures
\usepackage{minitoc} %table of contacts just for the supplementary
\noptcrule 

\usepackage[export]{adjustbox}
\usepackage[caption=false]{subfig}
\usepackage{hyperref}  % in case we want PDF bookmarks
\hypersetup{colorlinks=true, citecolor=blue, urlcolor=blue, linkcolor=blue}
\usepackage{dcolumn}   % needed for some tables
\usepackage{bm}        % for math
\usepackage{amssymb}   % for math
\usepackage{amsmath}   % for math also!
\usepackage{euscript}
\usepackage{verbatim}
\usepackage{setspace}
\usepackage{xcolor}
\usepackage{amsfonts}
\usepackage{braket}
\DeclareUnicodeCharacter{2212}{-}
\newcommand{\Hb}{4Hb-TaS$_2$}
\newcommand{\Hc}{$H_{c2}$}
\newcommand{\Csix}{$C_{6}$}
\newcommand{\Moire}{Moir{\'e}}
\newcommand{\be}{\begin{equation}}
\newcommand{\ee}{\end{equation}}
\newcommand{\bea}{\begin{eqnarray}}
\newcommand{\eea}{\end{eqnarray}}

\allowdisplaybreaks

\begin{document}

\title{Chiral to Nematic Crossover in the Superconducting State of \Hb{}}
\author{I. Silber}
%\email{itaisilber@mail.tau.ac.il}
\affiliation{School of Physics and Astronomy, Tel − Aviv University, Tel Aviv, 69978, Israel}
\author{S. Mathimalar}
\affiliation{Department of Condensed Matter Physics, Weizmann Institute of Science, Rehovot, Israel}
\author{I. Mangel}
\affiliation{Physics Department, Technion-Israel Institute of Technology, Haifa 32000, Israel}
\author{O. Green}
\affiliation{School of Physics and Astronomy, Tel − Aviv University, Tel Aviv, 69978, Israel}
\author{N. Avraham}
\affiliation{Department of Condensed Matter Physics, Weizmann Institute of Science, Rehovot, Israel}
\author{H. Beidenkopf}
\affiliation{Department of Condensed Matter Physics, Weizmann Institute of Science, Rehovot, Israel}
\author{I. Feldman}
\affiliation{Physics Department, Technion-Israel Institute of Technology, Haifa 32000, Israel}
\author{A. Kanigel}
\affiliation{Physics Department, Technion-Israel Institute of Technology, Haifa 32000, Israel}
\author{A. Klein}
\affiliation{Department of Physics, Faculty of Natural Sciences, Ariel University, Ariel 40700, Israel}
\affiliation{Department of Chemical Physics, The Weizmann Institute of Science, Rehovot 76100, Israel}
\author{M. Goldstein}
\affiliation{School of Physics and Astronomy, Tel − Aviv University, Tel Aviv, 69978, Israel}
\author{A. Banerjee}
\affiliation{Department of Physics, Ben-Gurion University of the Negev, Beer-Sheva 84105, Israel}
\author{E. Sela}
\affiliation{School of Physics and Astronomy, Tel − Aviv University, Tel Aviv, 69978, Israel}
\author{Y. Dagan}
\affiliation{School of Physics and Astronomy, Tel − Aviv University, Tel Aviv, 69978, Israel}

\begin{abstract}

\textbf{Most superconductors have an isotropic, single component order parameter and are well described by the standard (BCS) theory for superconductivity. Unconventional, multiple-component superconductors are exceptionally rare and are much less understood.
%The mechanisms responsible for unconventional superconductivity with multiple-component order parameters, and the possible resulting superconducting phase diagrams, are still elusive.
Here, we combine scanning tunneling microscopy and angle-resolved macroscopic transport for studying the candidate chiral superconductor, \Hb{}. We reveal quasi-periodic one-dimensional modulations in the tunneling conductance accompanied by two-fold symmetric superconducting critical-field. The strong modulation of the in-plane critical field, H$_{c2}$, points to a nematic, unconventional order parameter. However, the imaged vortex core is nearly circular symmetric, suggesting an isotropic order parameter. We reconcile this apparent discrepancy by modeling competition between a dominating chiral superconducting order parameter and a nematic one. The latter emerges close to the normal phase. Our results strongly support the existence of two-component superconductivity in \Hb{} and can provide valuable insights into other systems with coexistent charge order and superconductivity.}
\end{abstract}
\maketitle

Most superconductors have an isotropic order parameter, resulting from the short length scales of the charge screening and the electron-phonon interaction in metals and alloys. Unconventional superconductors, on the other hand, have an anisotropic and possible multiple-components order parameter, which transforms non-trivially under the symmetry operation of the underlying crystal \cite{Sigrist1991,Annett1995, Tsuei2000}. A unique situation emerges in chiral superconductors, when the order parameter is a time-reversal-symmetry-breaking linear combination of degenerate representations \cite{Kallin2016}. In practice, evidence for multi-component superconductivity has been detected in few compounds, e.g. NbSe$_2$ \cite{Hamill2021}, Bi$_2$Se$_3$ \cite{Pan2016,Du2017,Shen2017,Shingo2019,Cho2020}, and iron based superconductors \cite{Li2017,Kushnirenko2020}. Chiral superconductivity is even more elusive, with possible candidates being UPt$_3$ \cite{Strand2009,Schemm2014, Avers2020} and UTe$_2$ \cite{Metz2019,Ran2019,Jiao2020}.

Recently, mounting evidence points out that \Hb{} hosts a chiral superconducting state \cite{Ribak2020a,Nayak2021,Persky2022}. However, clear proof for the two-component superconducting state is lacking, and the superconducting phase diagram is yet to be determined.

\begin{figure*}[ht] %FIG1: most of the results: Structure, transitions, Hc2 -two fold, vortices, Stripes, theoretical phase diagram
  \captionsetup[subfigure]{labelformat=empty}
  \centering
  \subfloat[]{\raisebox{3.7ex}{\includegraphics[height=5.3cm]{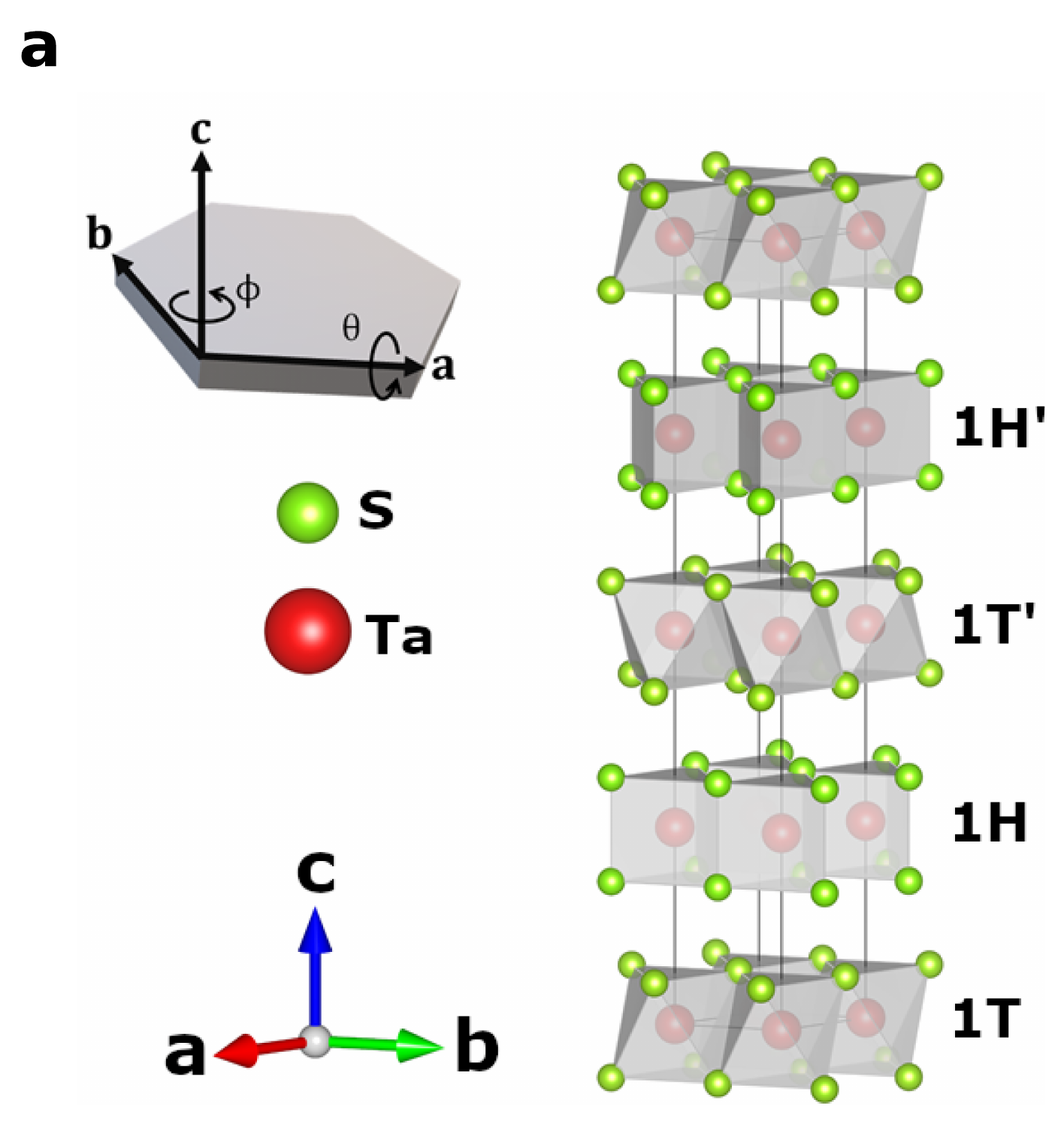}}\label{CrystalStracture}}
  \hspace{+4mm}
  \subfloat[]{\includegraphics[height=5.8cm]{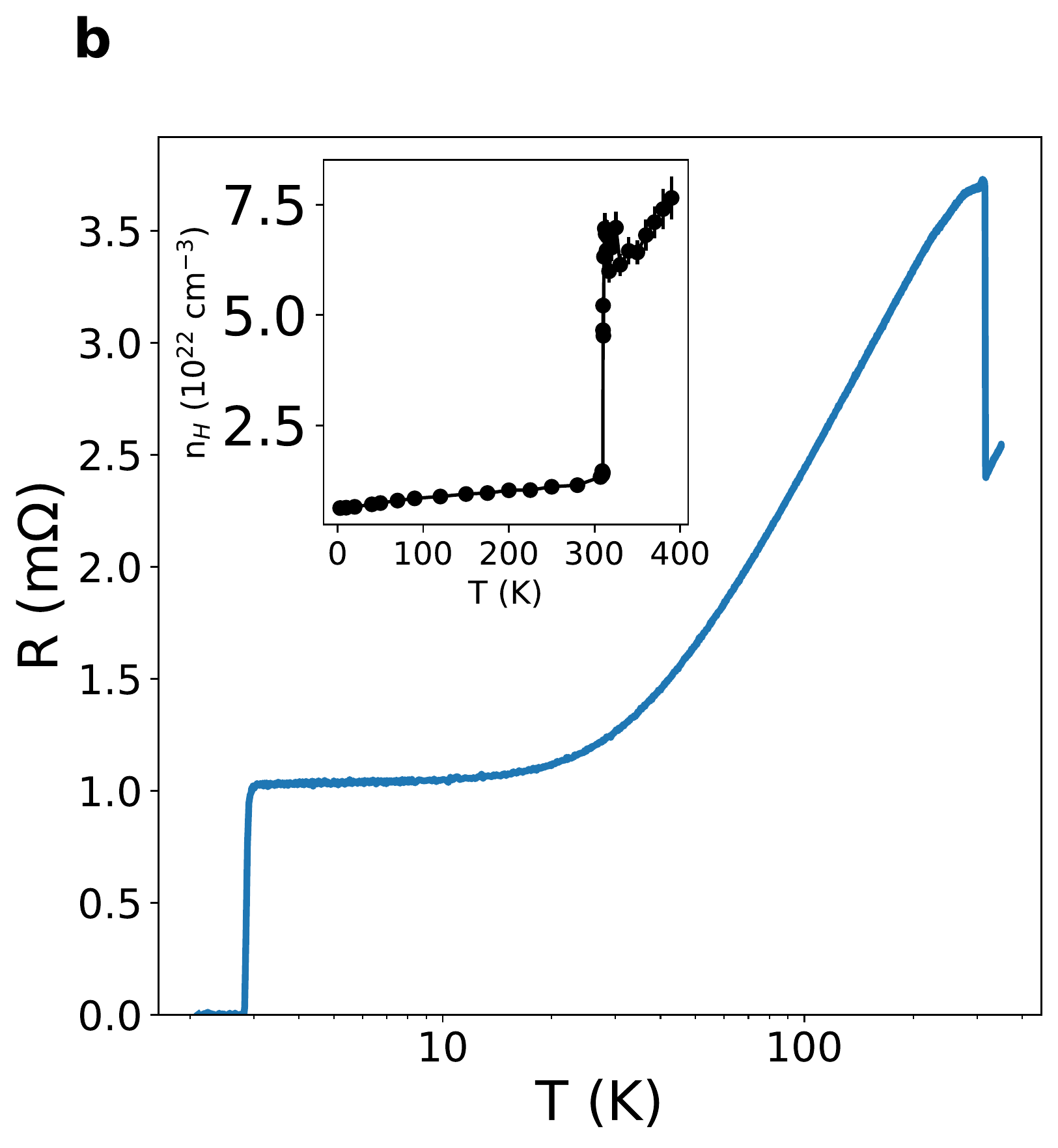}\label{RTGraph}}
  \hspace{+1.2mm}
  \subfloat[]{\raisebox{3.37ex}{\includegraphics[height=5.28cm]{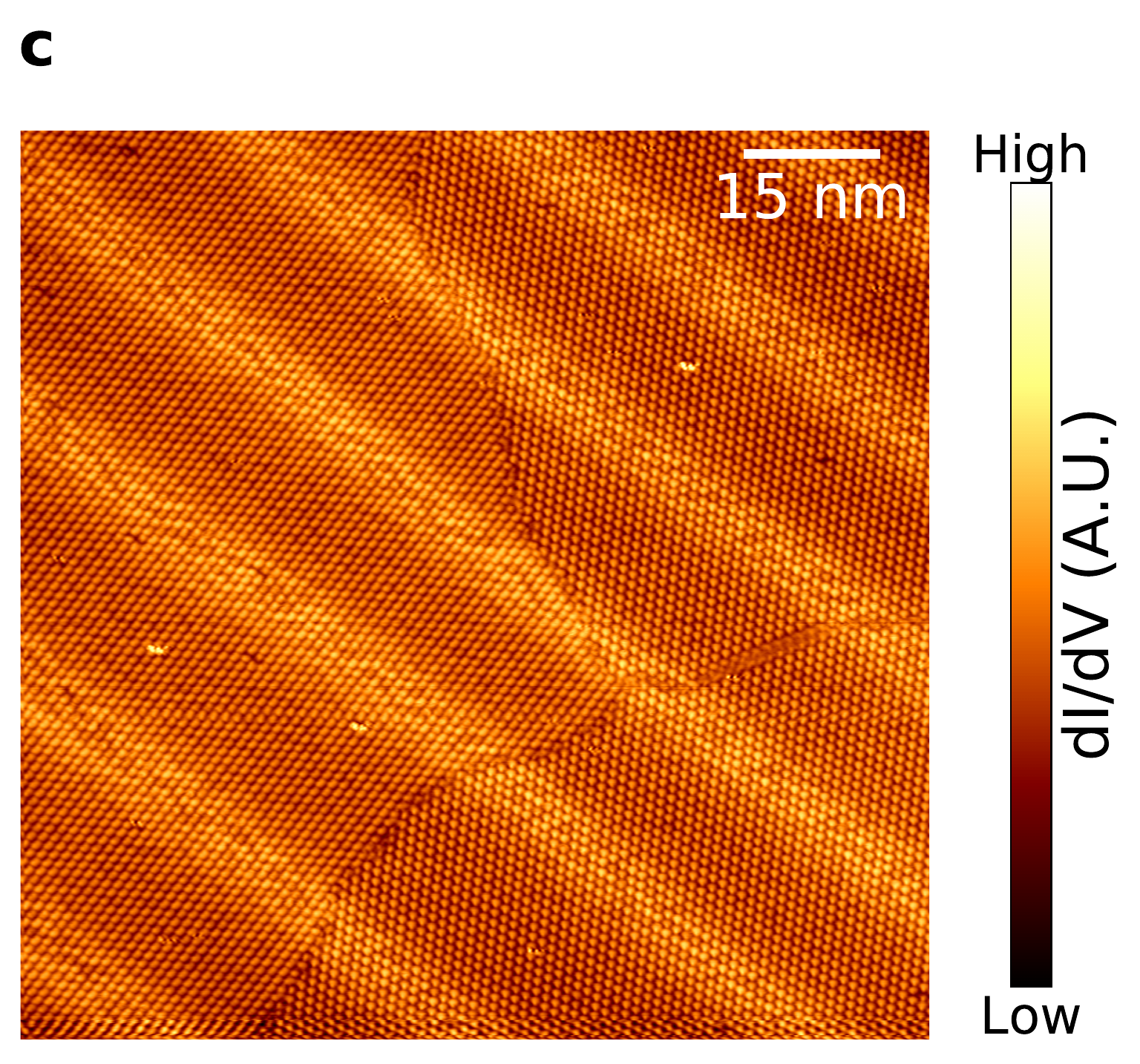}}\label{STMstripes}}
 \hfill
    %\subfloat[]{\includegraphics[height=5.8cm]{figures/HcPhi.pdf}\label{HcPhi}}
    \subfloat[]{\raisebox{-0.4ex}{\includegraphics[height=5.82cm]{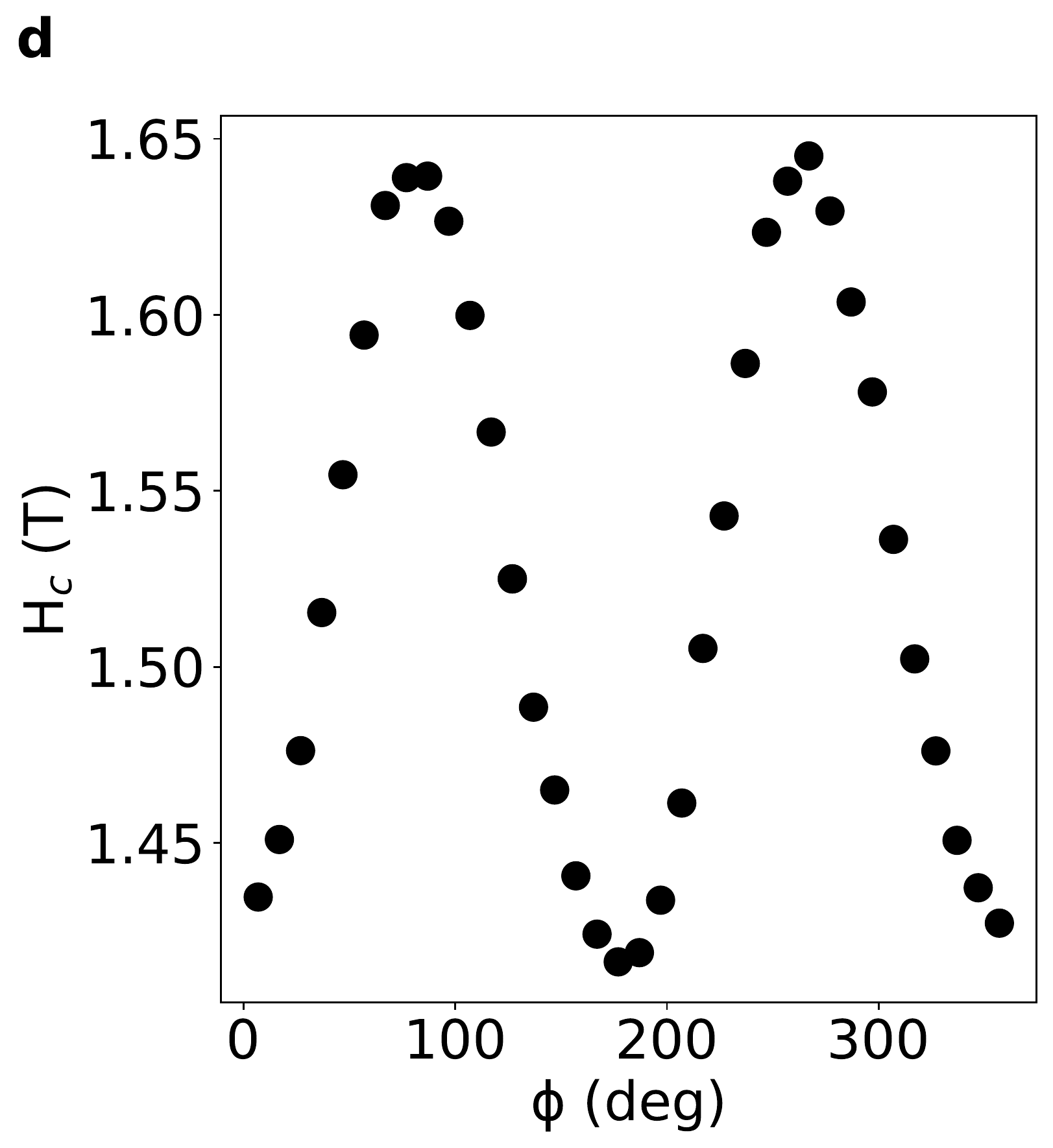}}\label{HcPhiFlake2halfK}}
   \subfloat[]{\raisebox{3.3ex}{\includegraphics[height=5.35cm]{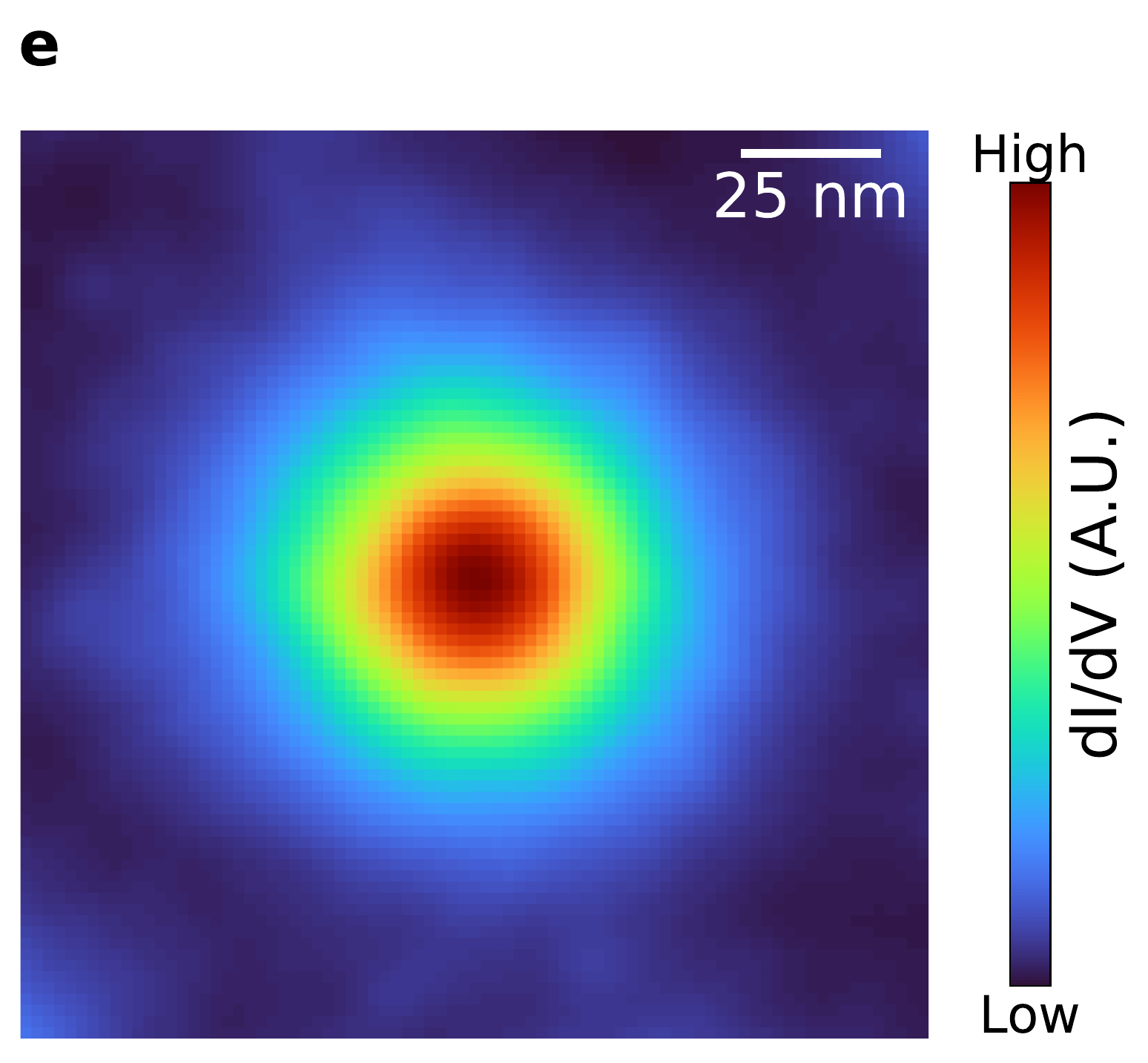}}\label{AveVortex}}
   \subfloat[]{\includegraphics[height=5.8cm]{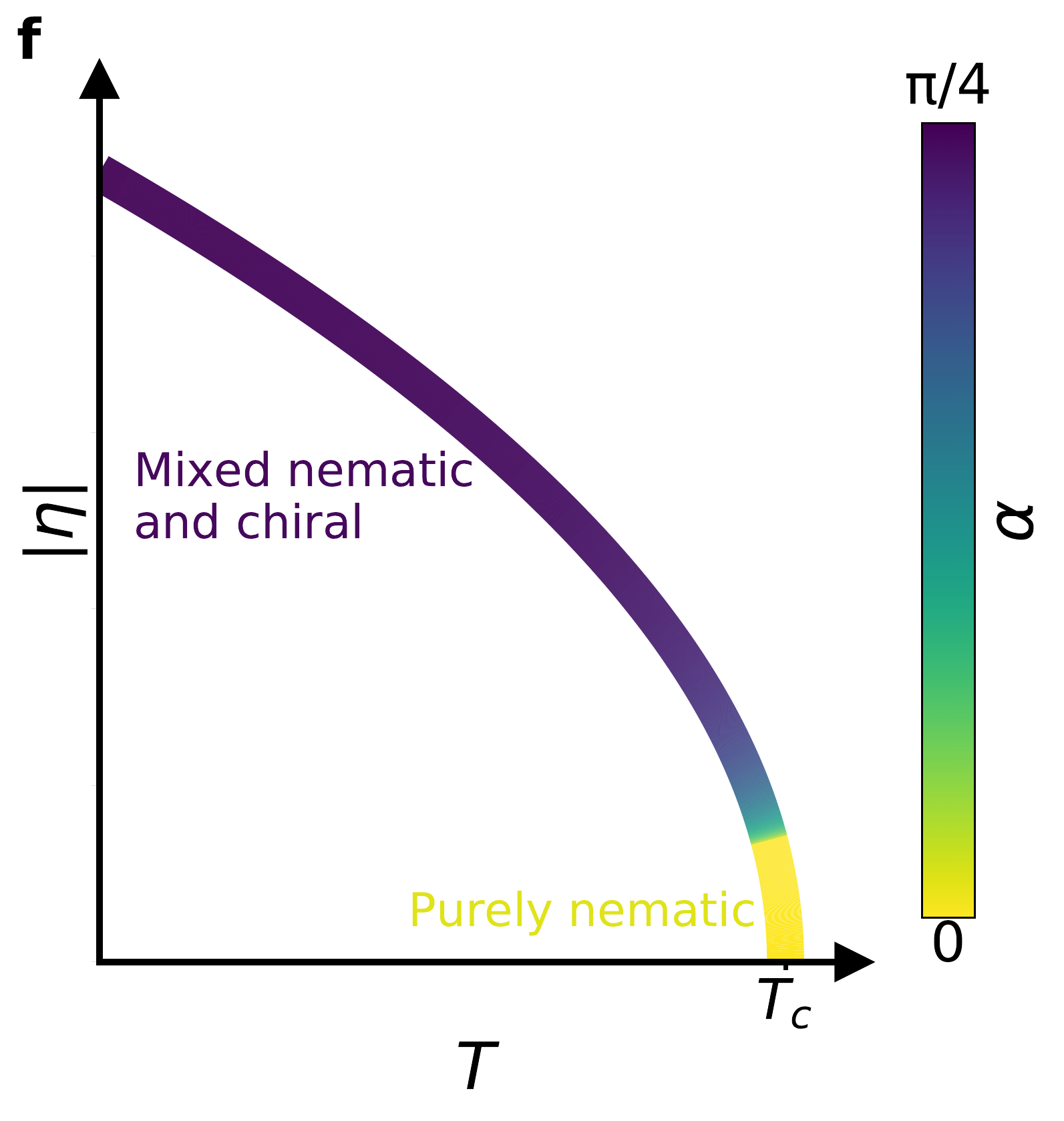}\label{Theory}}
\caption{Nematic and chiral superconductivity in \Hb{} and their origin. (a) Crystal structure of \Hb{} displaying alternately stacked quasi-2D layers of 1T-TaS$_2$ (1T) and half of the 2H-TaS$_2$ (1H) polymorphs. Inset: schematics of the hexagonal crystal, marking the crystal axes and the two directions of the field rotation in the experiment. (b) Resistance as a function of temperature. At $T=315$~K the charge density wave transition can be seen by the abrupt increase in the crystal resistance. The superconducting transition is observed at $T_c=2.7$~K. Note the logarithmic temperature scale. Inset: Hall number as a function of temperature. The charge density wave phase transition at $T=315$~K results in a strong reduction in Hall number. (c) STM topographic image, where each dot represents a single site in the $\sqrt13\times\sqrt13$ CDW patern. The stripes are clearly seen, and they extend across a CDW domain boundary. (d) The critical field (at $T=2.5$~K) has a clear 2-fold symmetry as a function of the in-plane direction of the applied magnetic field. (e) Average STM image of a vortex core. The density of states as a function of location, averaged over several vortices, is clearly isotropic. (f) The suggested phase diagram displays the variation of the order parameter $\eta$ with temperature. A nematic state is favored near $T_c$, whereas the order parameter becomes chiral at low temperatures. The nematic-chiral crossover is parametrized by the angle $\alpha$ %keeping the phase $\gamma = \frac{1}{2}$ 
(see details in the text).}\label{FIG1}
\end{figure*}

\Hb{} is a naturally occurring van der Waals heterostructure of alternating 1T- and 1H-TaS$_2$ layers (Fig.~\ref{CrystalStracture}). 1T-TaS$_2$ is considered to be a Mott insulator and a quantum spin liquid candidate \cite{Kratochvilova2017, Ribak2017, Law2017, Klanjsek2017}. 2H-TaS$_2$ is a superconductor with a critical temperature of $0.7$~K \cite{Bhoi2016}. \Hb{} undergoes a charge density wave transition at $T=315$~K \cite{Wilson1975}, which is clearly observed as an abrupt increase in the resistance (see Fig.~\ref{RTGraph}) accompanied by a five-fold reduction of the Hall number (see inset of Fig.~\ref{RTGraph} and Extended Data Fig.~\ref{HallAI9supp}). Nevertheless, the system remains metallic and undergoes a superconducting transition at $T_c=2.7$~K, as can be seen in Fig.~\ref{RTGraph}. 

We have performed scanning tunneling microscopy (STM) measurements. Each point in Fig.~\ref{STMstripes} is the 1T charge density wave (CDW) super-cell, which dominates over the weaker 1H charge density modulations~\cite{Nayak2021}. Surprisingly, we found modulations of the tunneling conductance in the form of highly oriented stripes with a separation of about $\approx 20 $ nm, see Fig.~\ref{STMstripes}. These stripes break the \Csix{} symmetry of the \Hb{} crystal.

Remarkably, this symmetry breaking is manifested in the superconducting state. When rotating the magnetic field in the basal plane, parallel to the TaS$_2$ planes of the sample, we observe significant two-fold modulations of the critical field \Hc{} (Fig.~\ref{HcPhiFlake2halfK}). We link the minimal critical field with the direction of the stripes as both are parallel to an in-plane crystal axis. The stripes found by our STM data allow for nematic symmetry breaking in \Hb{}, including its  superconducting state. Therefore, we interpret the two-fold critical field modulation as the signature of a multiple-component nematic superconducting order parameter, following Refs.~\cite{Fu2010,Fu2014,Venderbos2016,Hecker2018}.
%Following Ref.~\cite{Fu2010,Fu2014,Hecker2018} we interpret the critical field modulation as the signature  of a nematic superconducting order parameter allowed by the symmetry breaking found by our STM data. 
Naively, an anisotropic \Hc{} would result in an anisotropy of the vortex core~\cite{Bi2018}. However, the density of states in a vortex core is perfectly isotropic as seen in Fig.~\ref{AveVortex}. We reconcile the observed nematic state and the isotropic vortex using a theoretical analysis invoking a crossover from a chiral order parameter, which dominates at low temperatures, to a nematic one, appearing near T$_c$. The calculated zero field phase diagram is presented in Fig.~\ref{Theory}. Our model predicts a smaller anisotropy of \Hc{} at low temperatures, as observed in the experiment and discussed further below.   

\begin{figure*}[ht] %FIG2: STM analysis
  \captionsetup[subfigure]{labelformat=empty}
  \centering
%\subfloat[]{\includegraphics[height=5.8cm]{figures/CrystalDirections.png}\label{CrystalDir}}
  \subfloat[]{\includegraphics[height=5.8cm]{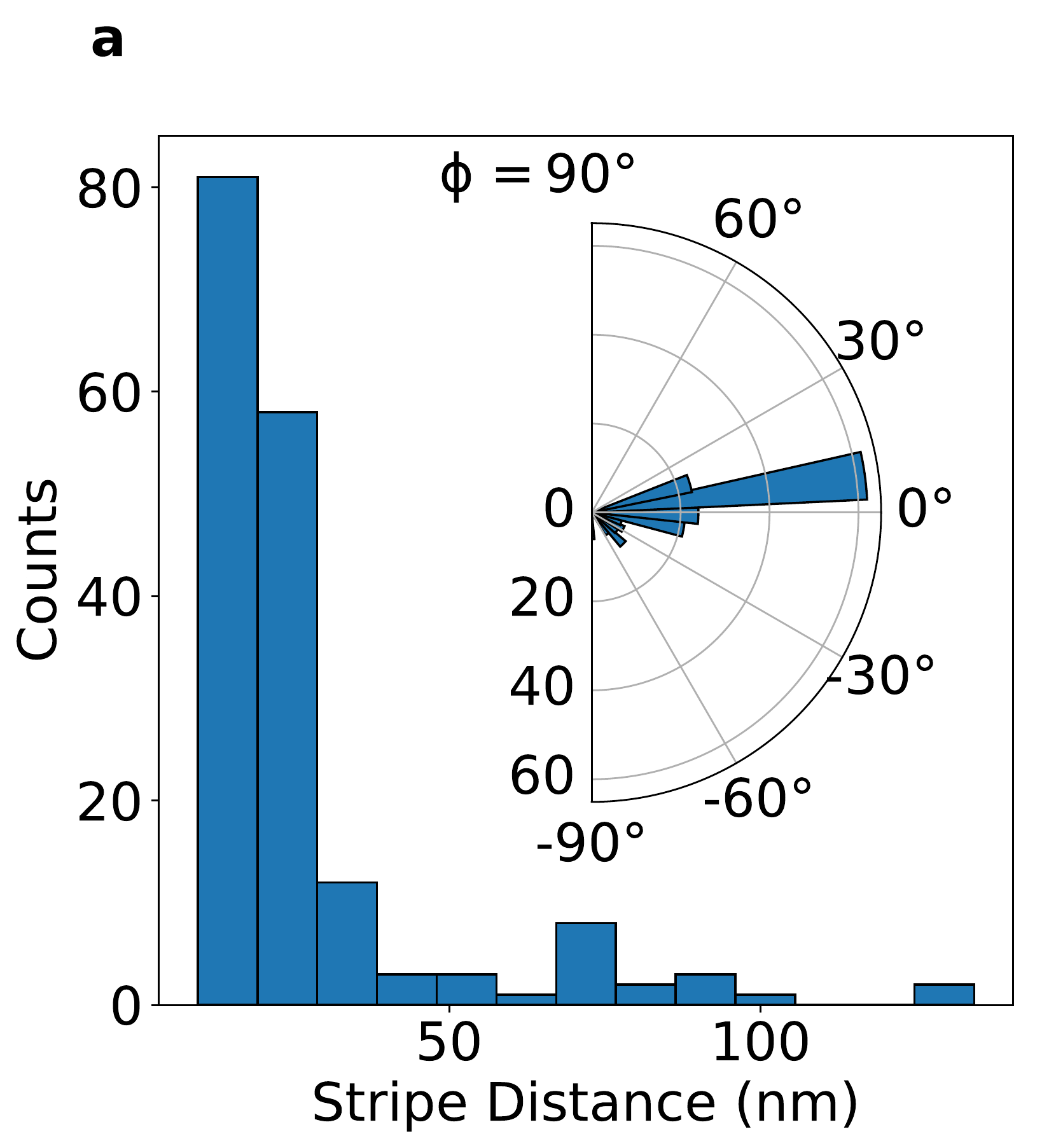}\label{StripeHistogram}}
  \subfloat[]{\raisebox{3.1ex}{\includegraphics[height=5.2cm]{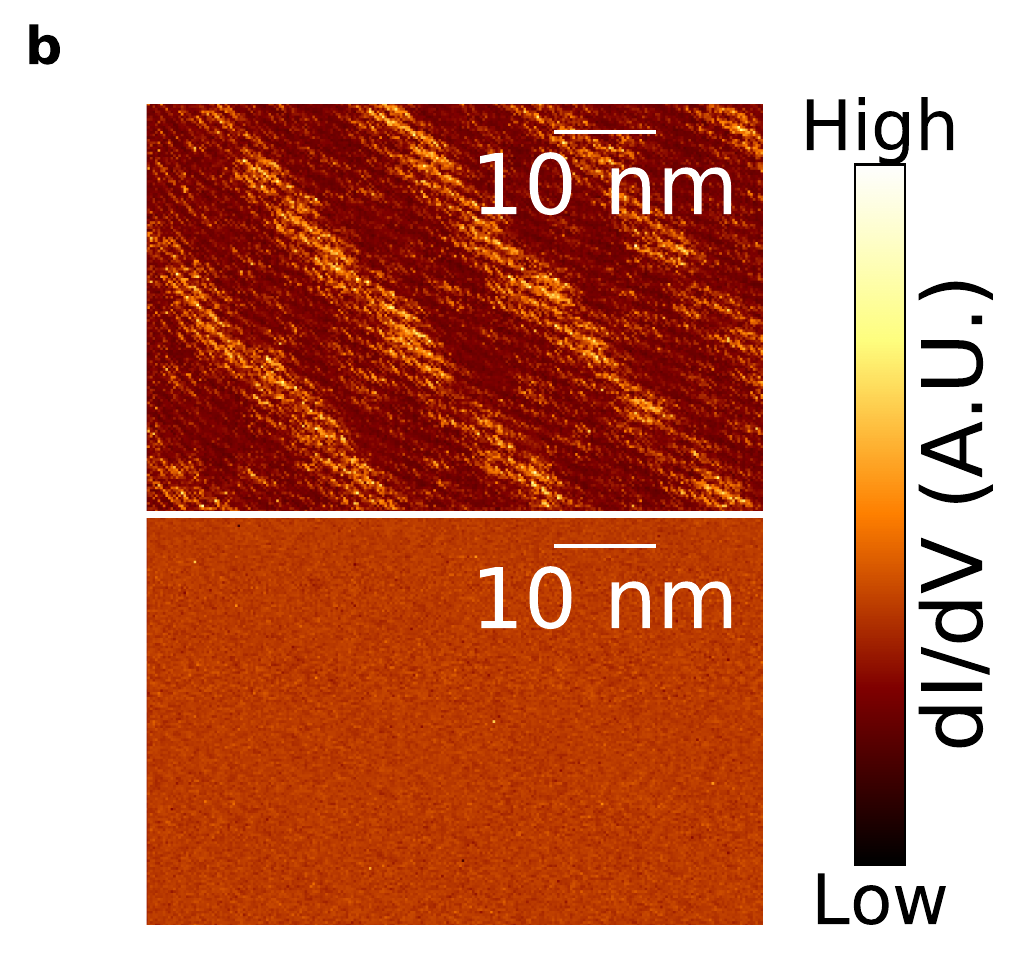}} \label{STMstripesdiffE}}
  \subfloat[]{\includegraphics[height=5.8cm]{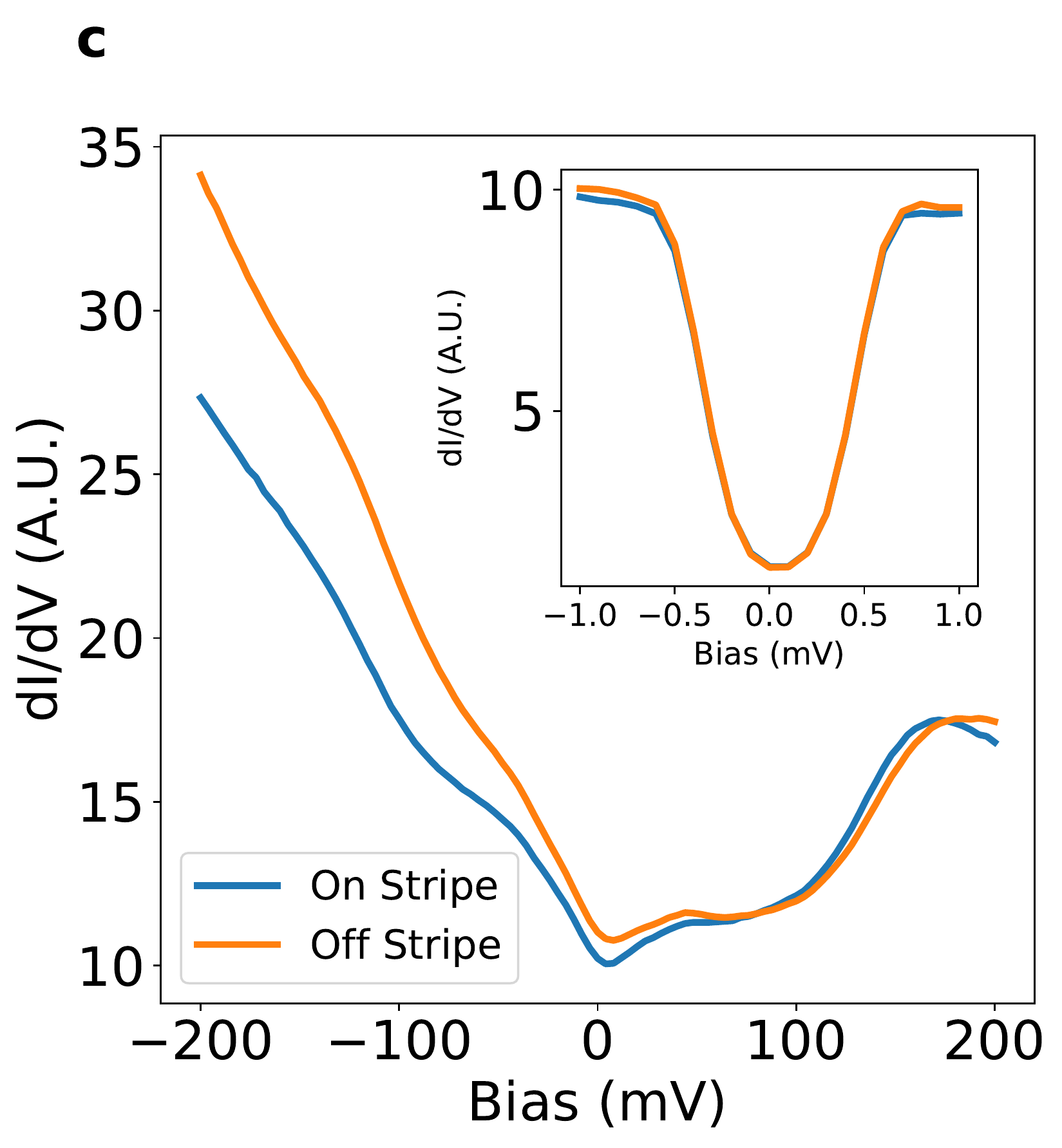} \label{STMdIdV}}
  %\subfloat[]{\includegraphics[height=5.8cm]{figures/STMvortices.pdf}\label{VortexProfile}}

\caption{Quasi-periodic conductance modulations in the form of  stripes in \Hb{}. (a) Distance and angular distributions of the stripes, showing a stripe separation with a median $P_{c} = 19.6$nm and a clear orientation, roughly parallel to a specific crystal axis. (b) Differential conductance maps of the same field of view measured at $E= 160$~meV (top), where the stripes are clearly seen, and at $E = -140$~meV (bottom), where the stripes are invisible. The absence of stripes at different energies shows that they are not simply a topographic modulation. (c) Spatially averaged dI/dV profiles measured on and off stripe, at $4$~K. Inset: Low-bias differential conductance spectra measured at $0.4$~K on and off stripe. The superconducting gap is unchanged, meaning the superconductivity is homogeneous throughout the sample. 
%(c) At low fields, the vortices are very isotropic, as can be seen from the spatial modulation of the density of states around a vortex core at various angels.
}\label{FIG2}
\end{figure*}

We now focus on the characterization of the stripes and their possible origin. The stripes are well oriented across the entire sample and define a specific direction over a macroscopic length scale, roughly parallel to a crystal axis, as seen in Fig.~\ref{StripeHistogram}. Their direction persists even when crossing a CDW domain wall, as demonstrated in Fig.~\ref{STMstripes}. We focus on the regions with abundant stripes and observe a quasi-periodic modulation of $\approx 19.6$~nm. In some rare regions no or only a single stripe is found, and the period is ill-defined.

We have performed local spectroscopic measurements on and off stripe, both in the normal and in the superconducting state well below $T_{c}$. We found that, for negative voltage biases, the density of states on a stripe differs from off a stripe (Figures~\ref{STMstripesdiffE} and ~\ref{STMdIdV}). This bias dependent difference suggests that the stripes represent a change in the local electronic density of states rather than just a topographic modulation. We further note that the superconducting gap itself, observed at low biases, remains the same on and off the stripes (Inset of Fig.~\ref{STMdIdV}). 

\begin{figure*}[ht] %FIG3: transport
  \captionsetup[subfigure]{labelformat=empty}
  \centering

  \subfloat[]{\includegraphics[height=5.89cm]{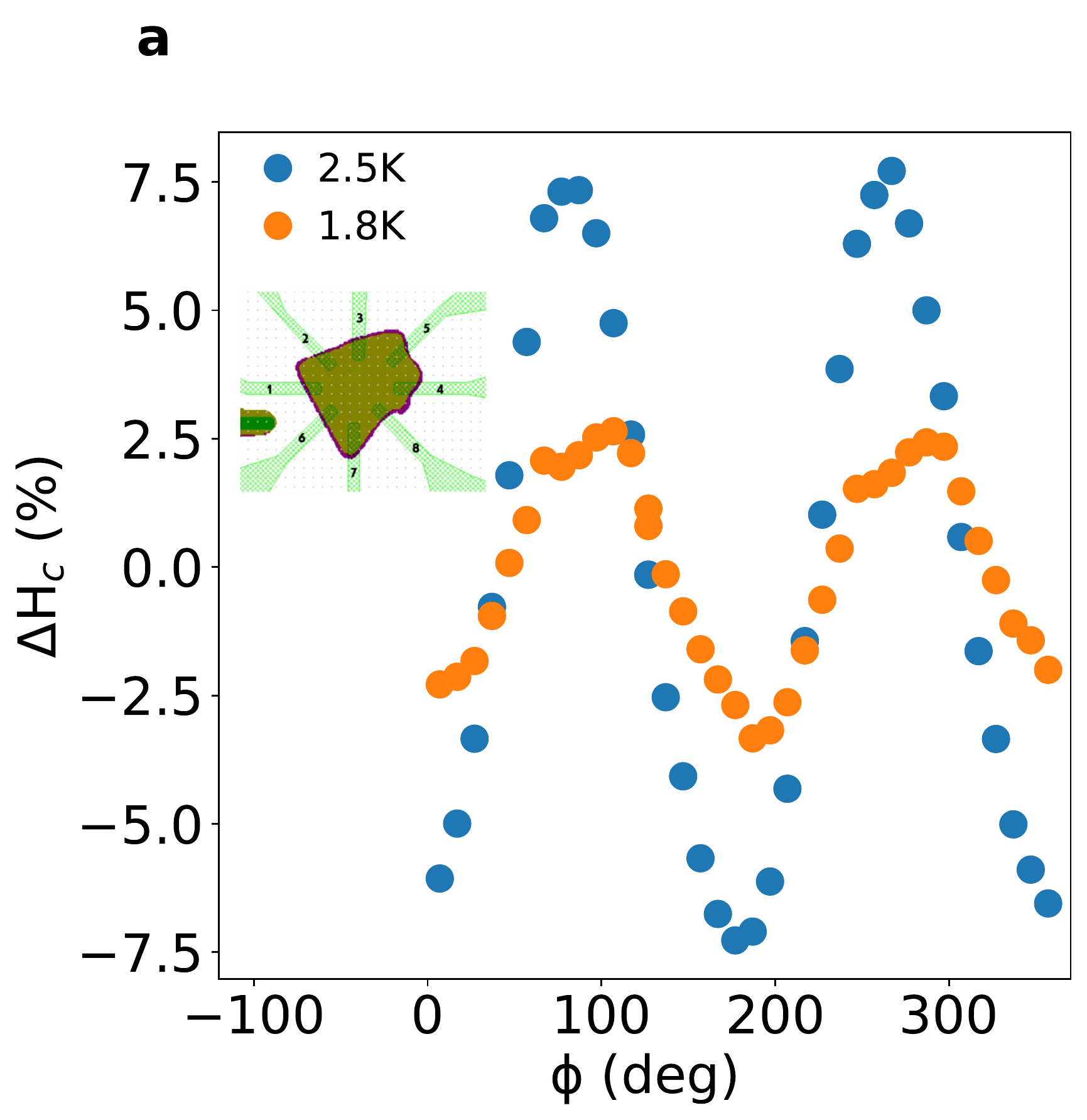}\label{HcPhiFlakediffT}}
  \subfloat[]{\includegraphics[height=5.8cm]{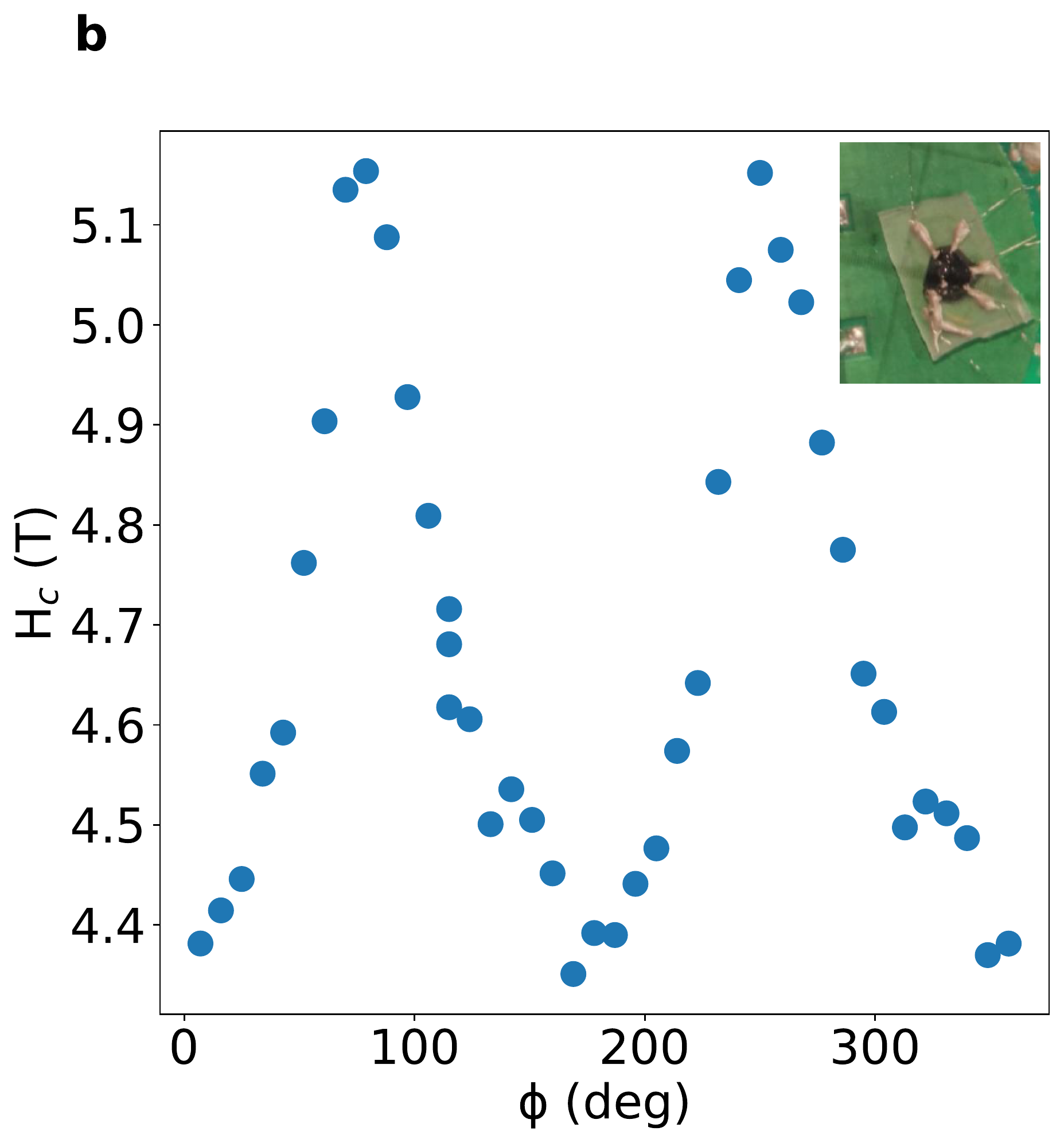}\label{HcPhi}}
  \subfloat[]{\includegraphics[height=5.8cm]{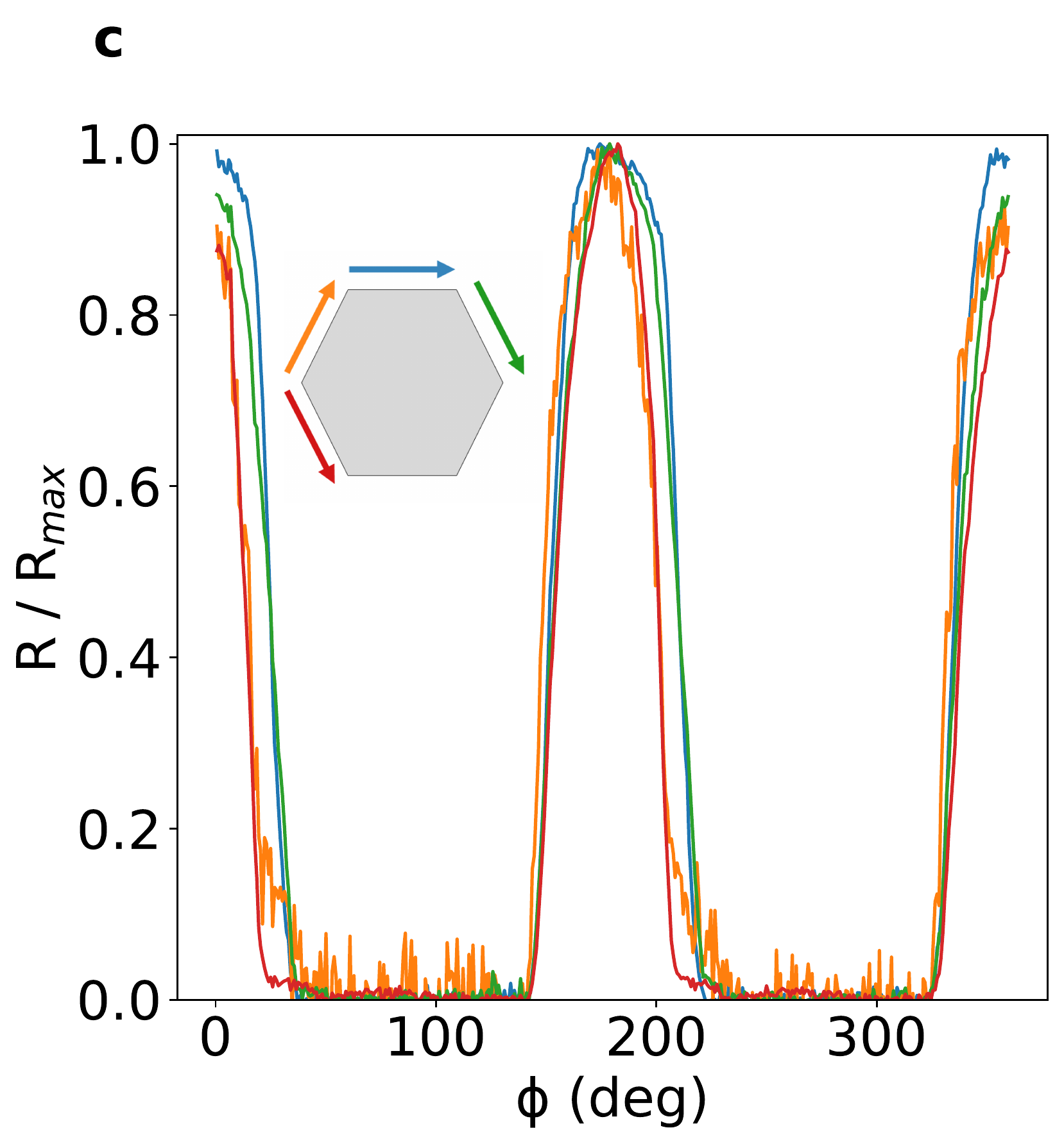}\label{RphiSC}}

    \caption{Nematic superconductivity in \Hb. (a) Critical fields of the flake sample at two temperatures. The critical field has a pronounced two-fold symmetry, whose anisotropy is larger at temperatures closer to the critical temperature. Inset: E-beam design of the eight gold contacts to the flake. (b) Similar measurement on a single crystal sample at $T=2.1K$, in which the crystal axes can be easily identified. The minimal critical field is observed when the field is applied parallel to an in-plane crystal axes. Inset: micrograph of the single-crystal sample with contacts attached. (c) Resistance measurements as a function of the applied magnetic field direction in the basal plane of the single crystal sample, for a constant magnetic field, $|H|=3.1T$, at $T=2.1K$. A peak in the resistance corresponds to a minimum in \Hc{} at these angles. The various colors represent four different current directions as depicted at the inset. Clearly, the overall behavior is independent of the current direction.}\label{FIG3}
\end{figure*}

\begin{figure}[ht] 
  \centering
   {\includegraphics[width=0.4\columnwidth]{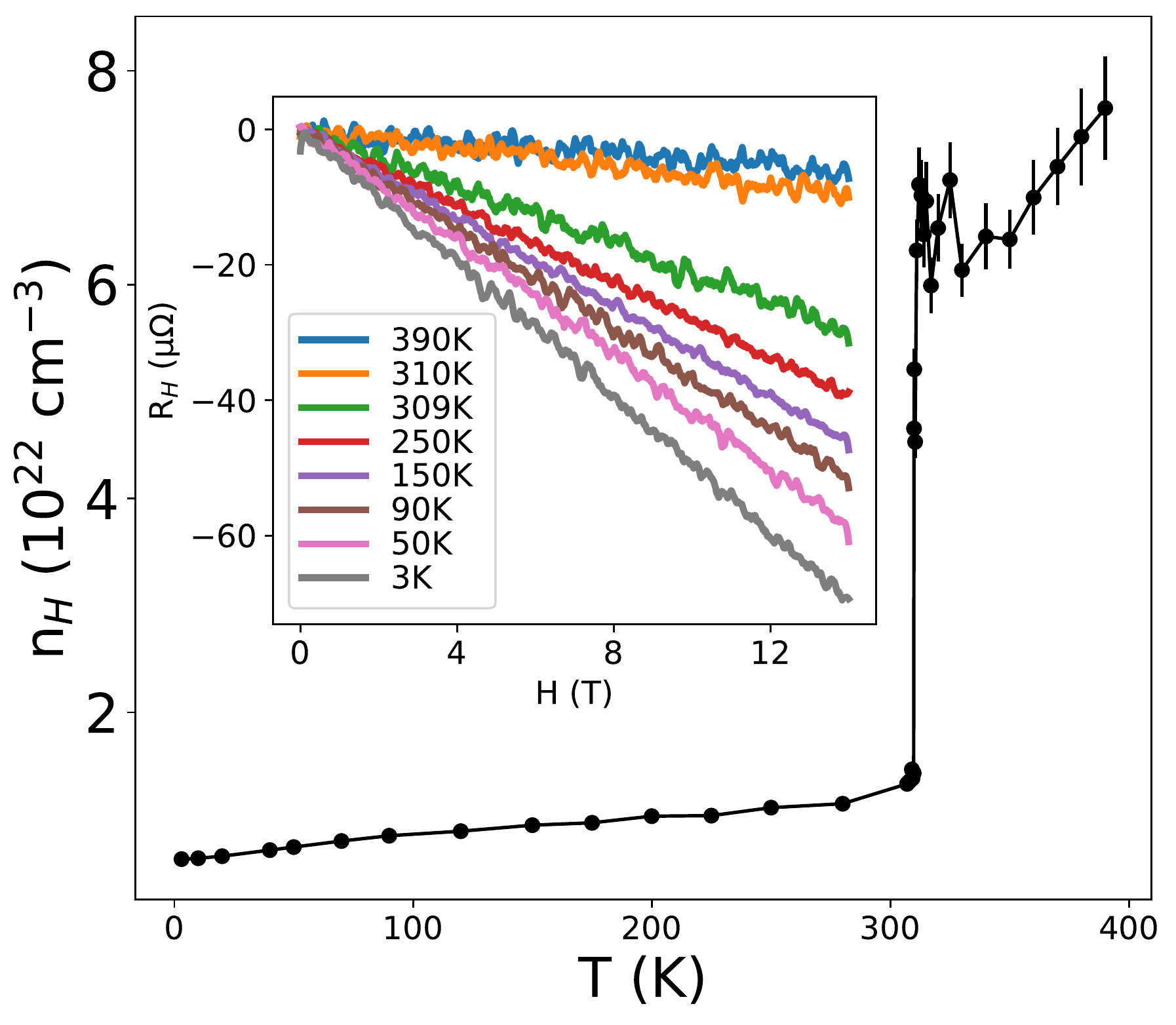}}

\caption{The charge density wave transition changes the carrier density dramatically. A careful measurement of the temperature dependence of the hall number reveals a decrease by a factor of five across the CDW transition. The Hall number is not simply related to the actual carriers density in TaS$_2$ polymorphs~\cite{Inada1980,Thompson1972} due to the complicated Fermi surfaces~\cite{Ribak2020a}. We interpret the dramatic change in the Hall number as a result of a Fermi surface reconstruction at the CDW transition that gaps major parts of Fermi surface. Inset: raw measurements of the Hall resistance as a function of the field at various temperatures.}
  \label{HallAI9supp}
\end{figure}

\begin{figure}[hb] 
  \centering
   {\includegraphics[width=0.4\columnwidth]{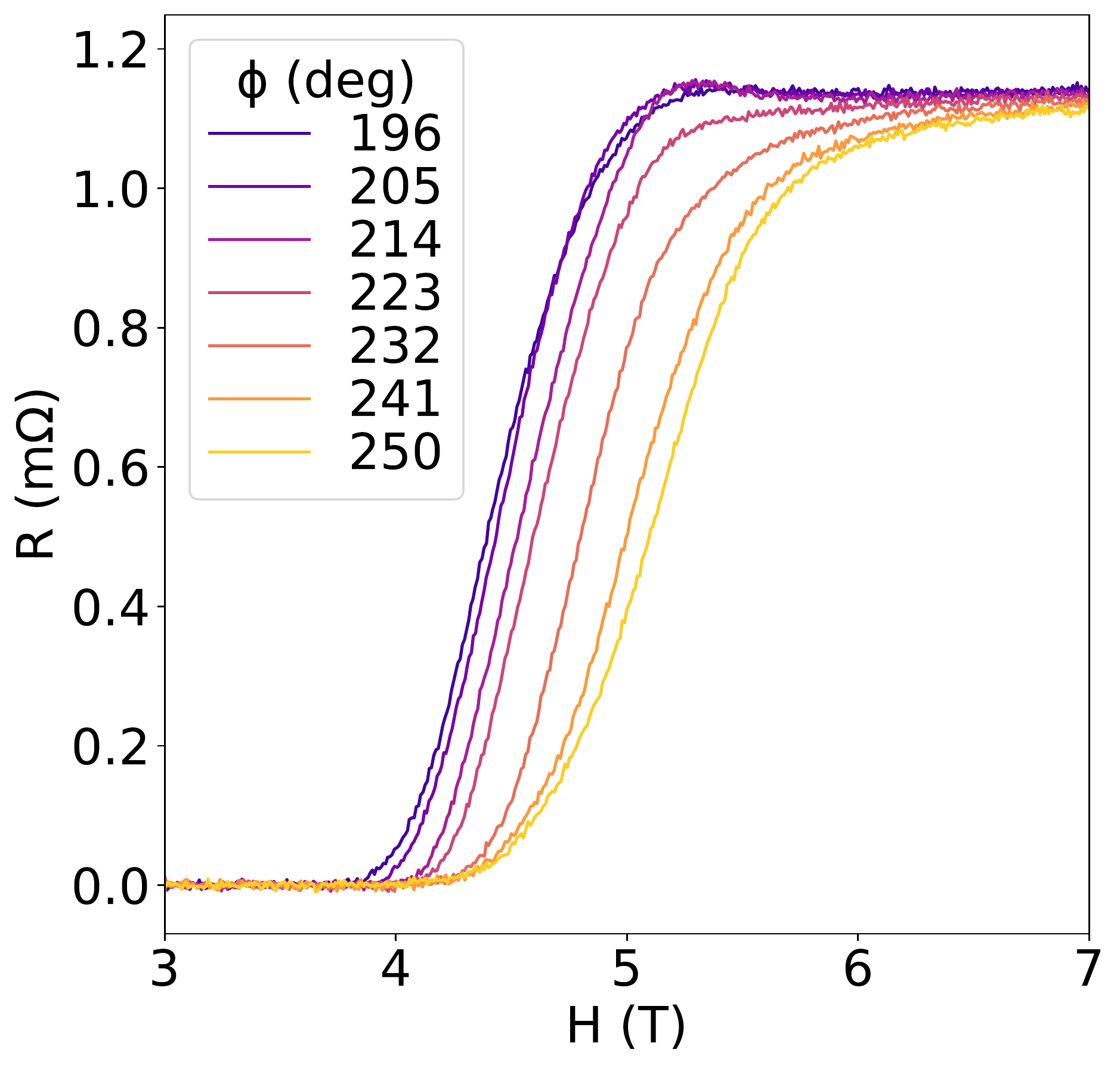}}

\caption{To deduce the critical field, magnetic field scans were performed at different in-plane angles while keeping a constant temperature. The critical field is clearly changing as a function of the in-plane angle. We show only a quarter cycle to avoid over-lapping of the measurements.}
  \label{RHdifferentPhi}
\end{figure}

As mentioned above, when we rotate the magnetic field in the basal plane, a remarkable two-fold symmetry of the critical magnetic field \Hc{} is detected, as seen in Fig.~\ref{FIG3}, and as detailed in Extended Data Fig.~\ref{RHdifferentPhi}. The minimum of \Hc{} is observed when the field is applied parallel to a crystal axis, similar to the direction of the stripes. Surprisingly, the anisotropy of the critical field is smaller as the temperature is decrease from $2.5$~K to $1.8$~K, and is much smaller at $T=0.4$~K, hardly resolved from the data (see Fig.~\ref{HcPhiTall}). 

One simple explanation for the anisotropy of \Hc{} is that it stems from a variation in the Josephson vortex pinning due to the stripes. To check this conjecture We model the stripes using a slow modulation of the inter-layer stacking configuration, yielding a variation of the perpendicular mass $m_z$ in a Ginzburg Landau (GL) theory along an in-plane coordinate, perpendicular to the observed stripes. Naturally, modulation of $m_z$ does not affect the round shape of the Abrikosov vortices in the plane. Solving the linearized GL equations, we generically find that the critical field is maximal parallel to the stripes. This is in contrast to our observations, ruling out this mechanism (see supplementary information for details). 
%ERAN: we just did perturbation theory but the modulation can be large! 
%the critical field modulations are usually small, contrast of our 20$\%$ modulations. Furthermore, 

Moreover, we have a unique situation in which closer to the critical temperature \Hb{} exhibits a clear nematic superconductivity, while at lower temperatures the superconducting state is much more isotropic (Fig. \ref{HcPhiTall}), as seen in the round vortices (Fig. \ref{AveVortex}) and the smaller variations in the in-plane critical field (Fig. \ref{HcPhiFlakediffT}). We also note that \Hb{} is different than most multi-component superconductors,  where the nematic behavior is observed throughout the entire temperature range \cite{Hamill2021,Pan2016,Du2017}.

Having excluded a conventional scenario, we now propose a competition between nematic and chiral superconductivity within a two-component GL theory \cite{Venderbos2016}. Such an explanation is consistent with the order parameter invoked by previous works \cite{Ribak2020a,Persky2022, Almoalem2022}.

The two-component order parameter can be written as $\vec{{\mathbf{\eta}}} =\eta \begin{pmatrix}
	\cos \alpha\\
	\sin \alpha ~ e^{i \gamma}
\end{pmatrix}
$,
where $\eta$ is the amplitude of the superconducting order parameter, and the angles $\alpha$ and $\gamma$ parametrize the nematic and chiral phases. In a purely nematic state $\gamma=0$, and  $\alpha$ dictates the direction of nematicity and resulting anisotropy. On the other hand, the purely chiral order, $\gamma \rightarrow \pm \pi/2$ and $\alpha=\pi/4$, is isotropic. Deep in the superconducting state, the system is described by a chiral order parameter~\cite{Ribak2020a}, leading to isotropic vortices. 

The origin of the emergent stripes is not fully understood. A possible scenario is that stacking two dimensional layers with a small mismatch and different CDW patterns generates a finite uniaxial strain (see supplementary material for more information). Regardless, we include this experimentally observed symmetry breaking via a $\epsilon_{xx}-\epsilon_{yy}$ term in the GL theory. Solving the GL equations for different temperatures discussed in the supplementary material, we obtain that $\gamma=\pi/2$, although $\alpha$ varies with $T$. Fig. \ref{Theory} shows that as the temperature increases and the order parameter is reduced, the strain term dominates to favor the nematic order near the critical temperature. 

Following this model, our results can be understood in the following way: at low temperatures, the order parameter is mostly chiral, in line with the isotropic vortex core in Fig. \ref{AveVortex} and the almost isotropic \Hc{} in Fig. \ref{HcPhiTall}. At temperatures closer to T$_c$, the order parameter is mostly nematic, consistent with the observed two-fold angular dependence of the critical field seen in Fig.~\ref{FIG3}. This theory predicts that vortices gradually become anisotropic at higher temperatures as the order parameter becomes nematic.

To summarize, we find a microscopic stripe pattern in \Hb{} and link it to the macroscopic two-fold symmetry of the superconducting critical field. The large variation in the critical field, of about 20\%, strongly supports the existence of a nematic order parameter close to the normal state. To reconcile theses findings with the indications of a chiral superconducting ground state, we offer a theoretical model that captures the crossover between a chiral order parameter to a nematic one. The latter is allowed by the strain field. We exclude a simpler explanation of Josephson-vortex pinning along the stripes that predicts a minimum in \Hc{} opposite to our findings. Our work indicates that the pairing in \Hb{} evolves from a nematic, time-reversal even state at high temperatures to a time-reversal breaking chiral order at very low temperatures, suggesting a unique superconducting phase diagram.

\section{Methods}
\textbf{Sample growth.} High-quality single crystals of \Hb{} were prepared using the chemical vapour transport method. Appropriate amounts of Ta and S were ground and mixed with a small amount of Se (1\% of the S amount). The powder was sealed in a quartz ampule, and a small amount of iodine was added as a transport agent. The ampule was placed in a three-zone furnace such that the powder is in the hot zone. After 30 days, single crystals with a typical size of 5.0 mm x 5.0 mm x 0.1 mm grew in the cold zone of the furnace. The fact that the van der Waals stacking in \Hb{} arrives in the form of a single crystal allows for clean, homogeneous and large samples.

\begin{figure}[ht] 
  \centering
   {\includegraphics[width=0.4\columnwidth]{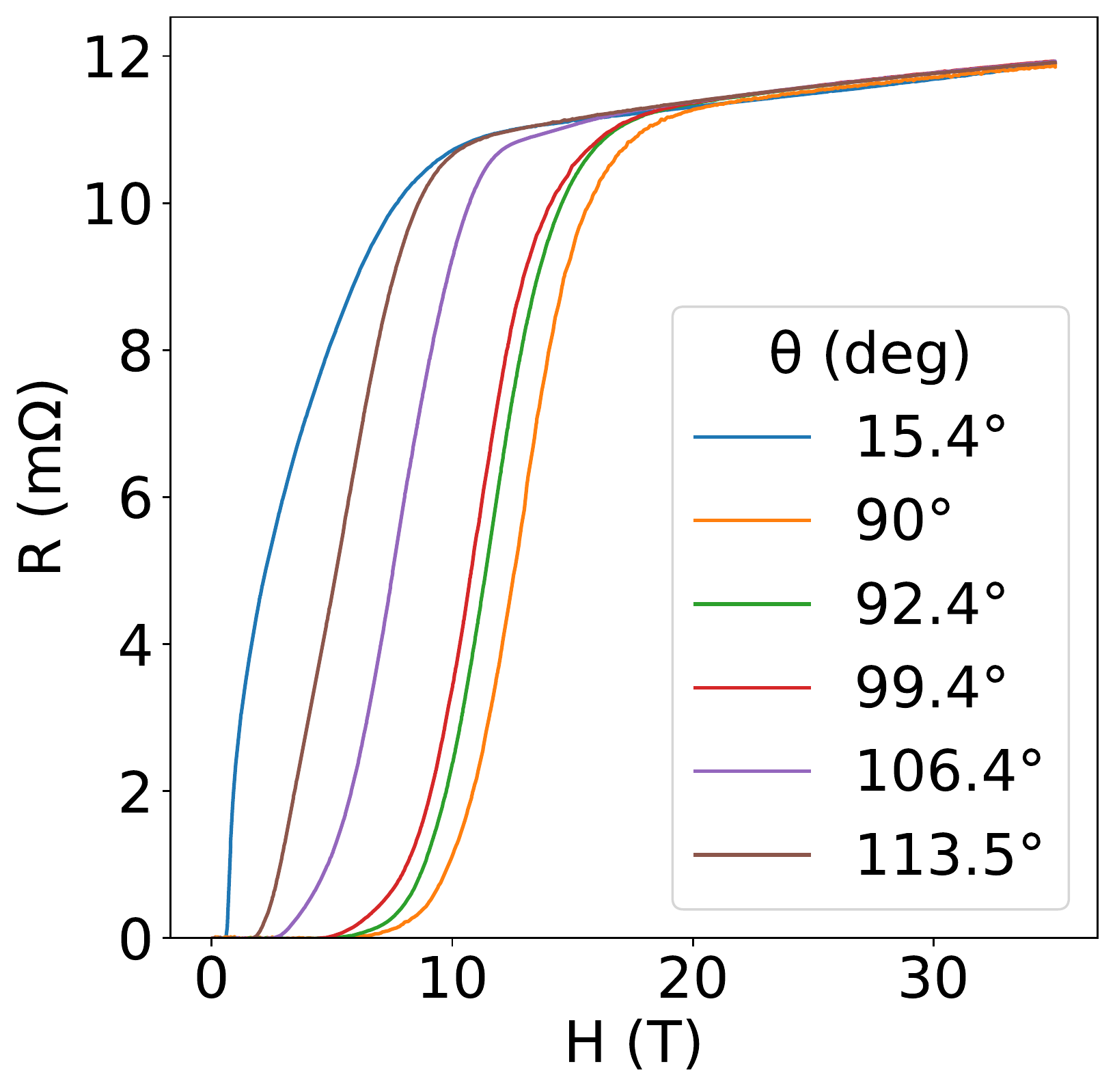}}

\caption{Field sweeps at different out-of-plane angles performed at $T=0.3$~K. When the applied magnetic field is rotated out of the basal plane ($\theta \neq 90^{\circ}$), the critical field is reduced drastically, as expected from the quasi-2D nature of the superconducting state. Remarkably, the normal state magneto-resistance is almost isotropic and non-saturating up to 33T.}
  \label{RHTall}
\end{figure}

\textbf{Transport Measurements}. Ohmic contacts were made by attaching platinum wires to the \Hb{} single crystals using silver epoxy, or by evaporating gold over the flake sample. Measurements up to 14T were taken in a cryogenic station with a mechanical rotator probe. The current, of the order of 1mA was supplied by a Keithley 6221 current source, and the voltage drop was measured using a Keithley 2182A nanovoltmeter. The flake sample was measured using a current of 250nA. High-field scans (Figs.~\ref{HcPhiTall} and \ref{RHTall}) were performed at the National High Magnetic Field Laboratory (NHMFL). In this case the resistance was measured using a Lakeshore 372 resistance bridge at low frequencies with a current of 0.316~mA. In-plane rotation was performed using a piezo stack, while monitoring the angle with three Hall sensors. The critical field was defined as the the field in which the resistance is half of the normal state resistance. For details about negating spurious wobbling effects and control experiments see the supplementary material section. To determine the Hall signal we averaged two current orientations and anti-symmetrized the results with respect to the applied magnetic field.

\textbf{STM measurements}. The samples were cleaved in STM load lock at ultra-high vacuum at room temperature. The cleaved crystals were then inserted directly to the 4K sample stage of the STM head for spectroscopic measurements. Commercial PtIr tips were treated on a freshly prepared Cu(111) single crystal for the stability of the tips and then used for the measurements. All the spectroscopic data were measured using standard lock-in techniques with frequency of 733Hz. In Fig.~\ref{STMstripes} the measurement was performed at a constant current mode with set value of 200~pA. We interpret the change at the signal as originating from a change in conductance, and not due to a change in topography, as the stripes' relative density of state change with bias (Fig.~\ref{STMstripesdiffE}). In Figs.~\ref{AveVortex}, \ref{STMdIdV} the voltage was set in the range of 0.1~mV to 10~mV depending on the bias scan range, and the AC modulation varied between 2~mV to 200~mV.

%\section{Acknowledgments}
\begin{acknowledgments}
 %\label{ACKNOWLEDGMENTS}
We thank M. Ben Shalom and J. Ruhman for useful discussion. We are indebted to David Graf for his valuable assistance at the high field laboratory. A part of the work presented here was performed at the National High Magnetic Field Laboratory (NHMFL), supported by the National Science Foundation Cooperative Agreement no. DMR-1644779, the State of Florida and the DOE. The experimental work performed at Tel Aviv university was supported by the Pazy Research Foundation Grant No. 326-1/22, Israel Science Foundation Grants No. 3079/20, the TAU Quantum Research Center and the Oren Family Chair for Experimental Physics.
M.G. was supported by the ISF and the Directorate for Defense Research and Development (DDR\&D) Grant No. 3427/21 and by the US-Israel Binational Science Foundation (BSF) Grant No. 2020072.
A.B. and E.S. acknowledge support from the European Research Council (ERC)  grant agreement No. 951541, ARO (W911NF-20-1-0013).
The experimental work performed at the Technion was supported by the ISF Grant No. 1263/21.

\end{acknowledgments}

\section{Author Contributions}
 \label{AUTHOR CONTRIBTUIONS}
I.F. and A.Kanigel prepared the \Hb{} single crystals. S.M., N.A and H.B. performed and analysed the STM measurements. M.G., A.B., E.S. and A.Klein proposed the theoretical explanation and performed the calculation of the phase diagrams, I.S. and O.G. performed the transport measurements over the single crystals. I.M. made and measured the flake sample.  Y.D. conceived the experiment. All authors discussed the findings, wrote the manuscript and reviewed its final version.

\bibliography{4HbTaS2bib} %same folder as tex file

\begin{thebibliography}{10}
\expandafter\ifx\csname url\endcsname\relax
  \def\url#1{\texttt{#1}}\fi
\expandafter\ifx\csname urlprefix\endcsname\relax\def\urlprefix{URL }\fi
\providecommand{\bibinfo}[2]{#2}
\providecommand{\eprint}[2][]{\url{#2}}

\bibitem{Sigrist1991}
\bibinfo{author}{Sigrist, M.} \& \bibinfo{author}{Ueda, K.}
\newblock \bibinfo{title}{Phenomenological theory of unconventional
  superconductivity}.
\newblock \emph{\bibinfo{journal}{Rev. Mod. Phys.}}
  \textbf{\bibinfo{volume}{63}}, \bibinfo{pages}{239--311}
  (\bibinfo{year}{1991}).
\newblock \urlprefix\url{https://link.aps.org/doi/10.1103/RevModPhys.63.239}.

\bibitem{Annett1995}
\bibinfo{author}{Annett, J.~F.}
\newblock \bibinfo{title}{Unconventional superconductivity}.
\newblock \emph{\bibinfo{journal}{Contemporary Physics}}
  \textbf{\bibinfo{volume}{36}}, \bibinfo{pages}{423--437}
  (\bibinfo{year}{1995}).

\bibitem{Tsuei2000}
\bibinfo{author}{Tsuei, C.~C.} \& \bibinfo{author}{Kirtley, J.~R.}
\newblock \bibinfo{title}{Pairing symmetry in cuprate superconductors}.
\newblock \emph{\bibinfo{journal}{Reviews of Modern Physics}}
  \textbf{\bibinfo{volume}{72}}, \bibinfo{pages}{969} (\bibinfo{year}{2000}).
\newblock
  \urlprefix\url{https://journals.aps.org/rmp/abstract/10.1103/RevModPhys.72.969}.

\bibitem{Kallin2016}
\bibinfo{author}{Kallin, C.} \& \bibinfo{author}{Berlinsky, J.}
\newblock \bibinfo{title}{Chiral superconductors}.
\newblock \emph{\bibinfo{journal}{Reports on Progress in Physics}}
  \textbf{\bibinfo{volume}{79}}, \bibinfo{pages}{054502}
  (\bibinfo{year}{2016}).
\newblock
  \urlprefix\url{https://iopscience.iop.org/article/10.1088/0034-4885/79/5/054502}.

\bibitem{Hamill2021}
\bibinfo{author}{Hamill, A.} \emph{et~al.}
\newblock \bibinfo{title}{Two-fold symmetric superconductivity in few-layer
  {NbSe$_2$}}.
\newblock \emph{\bibinfo{journal}{Nature Physics 2021 17:8}}
  \textbf{\bibinfo{volume}{17}}, \bibinfo{pages}{949--954}
  (\bibinfo{year}{2021}).
\newblock \urlprefix\url{https://www.nature.com/articles/s41567-021-01219-x}.

\bibitem{Pan2016}
\bibinfo{author}{Pan, Y.} \emph{et~al.}
\newblock \bibinfo{title}{Rotational symmetry breaking in the topological
  superconductor {Sr$_x$Bi$_2$Se$_3$} probed by upper-critical field
  experiments}.
\newblock \emph{\bibinfo{journal}{Scientific Reports}}
  \textbf{\bibinfo{volume}{6}}, \bibinfo{pages}{28632} (\bibinfo{year}{2016}).
\newblock \urlprefix\url{https://doi.org/10.1038/srep28632}.

\bibitem{Du2017}
\bibinfo{author}{Du, G.} \emph{et~al.}
\newblock \bibinfo{title}{Superconductivity with two-fold symmetry in
  topological superconductor {Sr$_x$Bi$_2$Se$_3$}}.
\newblock \emph{\bibinfo{journal}{Science China Physics, Mechanics \&
  Astronomy}} \textbf{\bibinfo{volume}{60}}, \bibinfo{pages}{037411}
  (\bibinfo{year}{2017}).
\newblock \urlprefix\url{https://doi.org/10.1007/s11433-016-0499-x}.

\bibitem{Shen2017}
\bibinfo{author}{Shen, J.} \emph{et~al.}
\newblock \bibinfo{title}{Nematic topological superconducting phase in nb-doped
  {Bi$_2$Se$_3$}}.
\newblock \emph{\bibinfo{journal}{npj Quantum Materials}}
  \textbf{\bibinfo{volume}{2}}, \bibinfo{pages}{59} (\bibinfo{year}{2017}).
\newblock \urlprefix\url{https://doi.org/10.1038/s41535-017-0064-1}.

\bibitem{Shingo2019}
\bibinfo{author}{Yonezawa, S.}
\newblock \bibinfo{title}{Nematic superconductivity in doped {Bi$_2$Se$_3$}
  topological superconductors}.
\newblock \emph{\bibinfo{journal}{Condensed Matter}}
  \textbf{\bibinfo{volume}{4}} (\bibinfo{year}{2019}).
\newblock \urlprefix\url{https://www.mdpi.com/2410-3896/4/1/2}.

\bibitem{Cho2020}
\bibinfo{author}{Cho, C.-w.} \emph{et~al.}
\newblock \bibinfo{title}{Z3-vestigial nematic order due to superconducting
  fluctuations in the doped topological insulators {Nb$_x$Bi$_2$Se$_3$} and
  {Cu$_x$Bi$_2$Se$_3$}}.
\newblock \emph{\bibinfo{journal}{Nature Communications}}
  \textbf{\bibinfo{volume}{11}}, \bibinfo{pages}{3056} (\bibinfo{year}{2020}).
\newblock \urlprefix\url{https://doi.org/10.1038/s41467-020-16871-9}.

\bibitem{Li2017}
\bibinfo{author}{Li, J.} \emph{et~al.}
\newblock \bibinfo{title}{Nematic superconducting state in iron pnictide
  superconductors}.
\newblock \emph{\bibinfo{journal}{Nature Communications}}
  \textbf{\bibinfo{volume}{8}}, \bibinfo{pages}{1880} (\bibinfo{year}{2017}).
\newblock \urlprefix\url{https://doi.org/10.1038/s41467-017-02016-y}.

\bibitem{Kushnirenko2020}
\bibinfo{author}{Kushnirenko, Y.~S.} \emph{et~al.}
\newblock \bibinfo{title}{Nematic superconductivity in {LiFeAs}}.
\newblock \emph{\bibinfo{journal}{Physical Review B}}
  \textbf{\bibinfo{volume}{102}}, \bibinfo{pages}{184502}
  (\bibinfo{year}{2020}).
\newblock \urlprefix\url{https://link.aps.org/doi/10.1103/PhysRevB.102.184502}.

\bibitem{Strand2009}
\bibinfo{author}{Strand, J.~D.}, \bibinfo{author}{Harlingen, D. J.~V.},
  \bibinfo{author}{Kycia, J.~B.} \& \bibinfo{author}{Halperin, W.~P.}
\newblock \bibinfo{title}{Evidence for complex superconducting order parameter
  symmetry in the low-temperature phase of {UPt$_3$} from josephson
  interferometry}.
\newblock \emph{\bibinfo{journal}{Physical Review Letters}}
  \textbf{\bibinfo{volume}{103}}, \bibinfo{pages}{197002}
  (\bibinfo{year}{2009}).
\newblock
  \urlprefix\url{https://journals.aps.org/prl/abstract/10.1103/PhysRevLett.103.197002}.

\bibitem{Schemm2014}
\bibinfo{author}{Schemm, E.~R.}, \bibinfo{author}{Gannon, W.~J.},
  \bibinfo{author}{Wishne, C.~M.}, \bibinfo{author}{Halperin, W.~P.} \&
  \bibinfo{author}{Kapitulnik, A.}
\newblock \bibinfo{title}{Observation of broken time-reversal symmetry in the
  heavy-fermion superconductor {UPt$_3$}}.
\newblock \emph{\bibinfo{journal}{Science}} \textbf{\bibinfo{volume}{345}},
  \bibinfo{pages}{190--193} (\bibinfo{year}{2014}).
\newblock \urlprefix\url{https://www.science.org/doi/10.1126/science.1248552}.

\bibitem{Avers2020}
\bibinfo{author}{Avers, K.~E.} \emph{et~al.}
\newblock \bibinfo{title}{Broken time-reversal symmetry in the topological
  superconductor {UPt$_3$}}.
\newblock \emph{\bibinfo{journal}{Nature Physics 2020 16:5}}
  \textbf{\bibinfo{volume}{16}}, \bibinfo{pages}{531--535}
  (\bibinfo{year}{2020}).
\newblock \urlprefix\url{https://www.nature.com/articles/s41567-020-0822-z}.

\bibitem{Metz2019}
\bibinfo{author}{Metz, T.} \emph{et~al.}
\newblock \bibinfo{title}{Point-node gap structure of the spin-triplet
  superconductor {UTe$_2$}}.
\newblock \emph{\bibinfo{journal}{Physical Review B}}
  \textbf{\bibinfo{volume}{100}}, \bibinfo{pages}{220504}
  (\bibinfo{year}{2019}).
\newblock
  \urlprefix\url{https://journals.aps.org/prb/abstract/10.1103/PhysRevB.100.220504}.

\bibitem{Ran2019}
\bibinfo{author}{Ran, S.} \emph{et~al.}
\newblock \bibinfo{title}{Nearly ferromagnetic spin-triplet superconductivity}.
\newblock \emph{\bibinfo{journal}{Science}} \textbf{\bibinfo{volume}{365}},
  \bibinfo{pages}{684--687} (\bibinfo{year}{2019}).
\newblock \urlprefix\url{https://www.science.org/doi/10.1126/science.aav8645}.

\bibitem{Jiao2020}
\bibinfo{author}{Jiao, L.} \emph{et~al.}
\newblock \bibinfo{title}{Chiral superconductivity in heavy-fermion metal
  {UTe$_2$}}.
\newblock \emph{\bibinfo{journal}{Nature}} \textbf{\bibinfo{volume}{579}},
  \bibinfo{pages}{523--527} (\bibinfo{year}{2020}).
\newblock \urlprefix\url{https://pubmed.ncbi.nlm.nih.gov/32214254/}.

\bibitem{Ribak2020a}
\bibinfo{author}{Ribak, A.} \emph{et~al.}
\newblock \bibinfo{title}{Chiral superconductivity in the alternate stacking
  compound {4Hb-TaS$_2$}}.
\newblock \emph{\bibinfo{journal}{Science Advances}}
  \textbf{\bibinfo{volume}{6}}, \bibinfo{pages}{9480--9507}
  (\bibinfo{year}{2020}).
\newblock
  \urlprefix\url{https://www.science.org/doi/abs/10.1126/sciadv.aax9480}.

\bibitem{Nayak2021}
\bibinfo{author}{Nayak, A.~K.} \emph{et~al.}
\newblock \bibinfo{title}{Evidence of topological boundary modes with
  topological nodal-point superconductivity}.
\newblock \emph{\bibinfo{journal}{Nature Physics 2021}} \bibinfo{pages}{1--7}
  (\bibinfo{year}{2021}).
\newblock \urlprefix\url{https://www.nature.com/articles/s41567-021-01376-z}.

\bibitem{Persky2022}
\bibinfo{author}{Persky, E.} \emph{et~al.}
\newblock \bibinfo{title}{Magnetic memory and spontaneous vortices in a van der
  waals superconductor}.
\newblock \emph{\bibinfo{journal}{Nature}} \textbf{\bibinfo{volume}{607}},
  \bibinfo{pages}{692--696} (\bibinfo{year}{2022}).
\newblock \urlprefix\url{https://www.nature.com/articles/s41586-022-04855-2}.

\bibitem{Kratochvilova2017}
\bibinfo{author}{Kratochvilova, M.} \emph{et~al.}
\newblock \bibinfo{title}{The low-temperature highly correlated quantum phase
  in the charge-density-wave {1T-TaS$_2$} compound}.
\newblock \emph{\bibinfo{journal}{npj Quantum Materials}}
  \textbf{\bibinfo{volume}{2}}, \bibinfo{pages}{42} (\bibinfo{year}{2017}).
\newblock \urlprefix\url{https://doi.org/10.1038/s41535-017-0048-1}.

\bibitem{Ribak2017}
\bibinfo{author}{Ribak, A.} \emph{et~al.}
\newblock \bibinfo{title}{Gapless excitations in the ground state of
  {1T-TaS$_2$}}.
\newblock \emph{\bibinfo{journal}{Physical Review B}}
  \textbf{\bibinfo{volume}{96}}, \bibinfo{pages}{195131}
  (\bibinfo{year}{2017}).
\newblock \urlprefix\url{https://link.aps.org/doi/10.1103/PhysRevB.96.195131}.

\bibitem{Law2017}
\bibinfo{author}{Law, K.~T.} \& \bibinfo{author}{Lee, P.~A.}
\newblock \bibinfo{title}{{1T-TaS$_2$} as a quantum spin liquid}.
\newblock \emph{\bibinfo{journal}{Proceedings of the National Academy of
  Sciences of the United States of America}} \textbf{\bibinfo{volume}{114}},
  \bibinfo{pages}{6996--7000} (\bibinfo{year}{2017}).
\newblock \urlprefix\url{https://www.pnas.org/content/114/27/6996
  https://www.pnas.org/content/114/27/6996.abstract}.

\bibitem{Klanjsek2017}
\bibinfo{author}{Klanjšek, M.} \emph{et~al.}
\newblock \bibinfo{title}{A high-temperature quantum spin liquid with polaron
  spins}.
\newblock \emph{\bibinfo{journal}{Nature Physics}}
  \textbf{\bibinfo{volume}{13}}, \bibinfo{pages}{1130} (\bibinfo{year}{2017}).
\newblock \urlprefix\url{https://doi.org/10.1038/nphys4212
  http://10.0.4.14/nphys4212
  https://www.nature.com/articles/nphys4212#supplementary-information}.

\bibitem{Bhoi2016}
\bibinfo{author}{Bhoi, D.} \emph{et~al.}
\newblock \bibinfo{title}{Interplay of charge density wave and multiband
  superconductivity in {2H-Pd$_x$TaSe$_2$}}.
\newblock \emph{\bibinfo{journal}{Scientific Reports 2016 6:1}}
  \textbf{\bibinfo{volume}{6}}, \bibinfo{pages}{1--10} (\bibinfo{year}{2016}).
\newblock \urlprefix\url{https://www.nature.com/articles/srep24068}.

\bibitem{Wilson1975}
\bibinfo{author}{Wilson, J.~A.}, \bibinfo{author}{Salvo, F. J.~D.} \&
  \bibinfo{author}{Mahajan, S.}
\newblock \bibinfo{title}{Charge-density waves and superlattices in the
  metallic layered transition metal dichalcogenides}.
\newblock \emph{\bibinfo{journal}{Advances in Physics}}
  \textbf{\bibinfo{volume}{24}}, \bibinfo{pages}{117--201}
  (\bibinfo{year}{1975}).
\newblock
  \urlprefix\url{https://www.tandfonline.com/doi/abs/10.1080/00018737500101391}.

\bibitem{Fu2010}
\bibinfo{author}{Fu, L.} \& \bibinfo{author}{Berg, E.}
\newblock \bibinfo{title}{Odd-parity topological superconductors: Theory and
  application to ${\mathrm{cu}}_{x}{\mathrm{bi}}_{2}{\mathrm{se}}_{3}$}.
\newblock \emph{\bibinfo{journal}{Phys. Rev. Lett.}}
  \textbf{\bibinfo{volume}{105}}, \bibinfo{pages}{097001}
  (\bibinfo{year}{2010}).
\newblock
  \urlprefix\url{https://link.aps.org/doi/10.1103/PhysRevLett.105.097001}.

\bibitem{Fu2014}
\bibinfo{author}{Fu, L.}
\newblock \bibinfo{title}{Odd-parity topological superconductor with nematic
  order: Application to ${\mathrm{cu}}_{x}{\mathrm{bi}}_{2}{\mathrm{se}}_{3}$}.
\newblock \emph{\bibinfo{journal}{Phys. Rev. B}} \textbf{\bibinfo{volume}{90}},
  \bibinfo{pages}{100509} (\bibinfo{year}{2014}).
\newblock \urlprefix\url{https://link.aps.org/doi/10.1103/PhysRevB.90.100509}.

\bibitem{Venderbos2016}
\bibinfo{author}{Venderbos, J. W.~F.}, \bibinfo{author}{Kozii, V.} \&
  \bibinfo{author}{Fu, L.}
\newblock \bibinfo{title}{Identification of nematic superconductivity from the
  upper critical field}.
\newblock \emph{\bibinfo{journal}{Physical Review B}}
  \textbf{\bibinfo{volume}{94}}, \bibinfo{pages}{94522} (\bibinfo{year}{2016}).
\newblock \urlprefix\url{https://link.aps.org/doi/10.1103/PhysRevB.94.094522}.

\bibitem{Hecker2018}
\bibinfo{author}{Hecker, M.} \& \bibinfo{author}{Schmalian, J.}
\newblock \bibinfo{title}{Vestigial nematic order and superconductivity in the
  doped topological insulator cuxbi2se3}.
\newblock \emph{\bibinfo{journal}{npj Quantum Materials}}
  \textbf{\bibinfo{volume}{3}}, \bibinfo{pages}{26} (\bibinfo{year}{2018}).
\newblock \urlprefix\url{https://doi.org/10.1038/s41535-018-0098-z}.

\bibitem{Bi2018}
\bibinfo{author}{Bi, C.} \emph{et~al.}
\newblock \bibinfo{title}{Direct visualization of the nematic superconductivity
  in {Cu$_x$Bi$_2$Se$_3$}}.
\newblock \emph{\bibinfo{journal}{Phys. Rev. X}} \textbf{\bibinfo{volume}{8}},
  \bibinfo{pages}{041024} (\bibinfo{year}{2018}).

\bibitem{Almoalem2022}
\bibinfo{author}{Almoalem, A.}
\newblock \bibinfo{title}{Unpublished literature}.
\newblock \emph{\bibinfo{journal}{unpublished}}  (\bibinfo{year}{2022}).

\bibitem{Inada1980}
\bibinfo{author}{Inada, R.}, \bibinfo{author}{Onuki, Y.} \&
  \bibinfo{author}{Tanuma, S.}
\newblock \bibinfo{title}{Hall effect of {1T-TaS$_2$} and {1T-TaSe$_2$}}.
\newblock \emph{\bibinfo{journal}{Physica B+C}} \textbf{\bibinfo{volume}{99}},
  \bibinfo{pages}{188--192} (\bibinfo{year}{1980}).

\bibitem{Thompson1972}
\bibinfo{author}{Thompson, A.~H.}, \bibinfo{author}{Gamble, F.~R.} \&
  \bibinfo{author}{Koehler, R.~F.}
\newblock \bibinfo{title}{Effects of intercalation on electron transport in
  tantalum disulfide}.
\newblock \emph{\bibinfo{journal}{Physical Review B}}
  \textbf{\bibinfo{volume}{5}}, \bibinfo{pages}{2811--2816}
  (\bibinfo{year}{1972}).

\bibitem{Conroy1972}
\bibinfo{author}{Conroy, L.~E.} \& \bibinfo{author}{Pisharody, K.~R.}
\newblock \bibinfo{title}{The preparation and properties of single crystals of
  the 1s and 2s polymorphs of tantalum disulfide}.
\newblock \emph{\bibinfo{journal}{Journal of Solid State Chemistry}}
  \textbf{\bibinfo{volume}{4}}, \bibinfo{pages}{345--350}
  (\bibinfo{year}{1972}).

\bibitem{Narayan1976}
\bibinfo{author}{Narayan, J.}
\newblock \bibinfo{title}{Coexistence of two charge density waves of different
  symmetry in transition metal dichalcogenide {4Hb-TaS$_2$}}.
\newblock \emph{\bibinfo{journal}{Applied Physics Letters}}
  \textbf{\bibinfo{volume}{29}}, \bibinfo{pages}{223--224}
  (\bibinfo{year}{1976}).
\newblock \urlprefix\url{https://doi.org/10.1063/1.89043}.

\bibitem{Scholz1982}
\bibinfo{author}{Scholz, G.~A.}, \bibinfo{author}{Singh, O.},
  \bibinfo{author}{Frindt, R.~F.} \& \bibinfo{author}{Curzon, A.~E.}
\newblock \bibinfo{title}{Charge density wave commensurability in {2H-TaS$_2$}
  and {Ag$_x$TaS$_2$}}.
\newblock \emph{\bibinfo{journal}{Solid State Communications}}
  \textbf{\bibinfo{volume}{44}}, \bibinfo{pages}{1455--1459}
  (\bibinfo{year}{1982}).

\bibitem{Ravnik2019}
\bibinfo{author}{Ravnik, J.} \emph{et~al.}
\newblock \bibinfo{title}{Strain-induced metastable topological networks in
  laser-fabricated {TaS$_2$} polytype heterostructures for nanoscale devices}.
\newblock \emph{\bibinfo{journal}{ACS Applied Nano Materials}}
  \textbf{\bibinfo{volume}{2}}, \bibinfo{pages}{3743--3751}
  (\bibinfo{year}{2019}).
\newblock \urlprefix\url{https://pubs.acs.org/doi/10.1021/acsanm.9b00644}.

\bibitem{Varga}
\bibinfo{author}{Varga, K.} \& \bibinfo{author}{Driscoll, J.~A.}
\newblock \emph{\bibinfo{title}{Computational nanoscience: Applications for
  molecules, clusters, and solids}} (\bibinfo{publisher}{Cambridge University
  Press}, \bibinfo{year}{2011}).

\bibitem{Etter2018}
\bibinfo{author}{Etter, S.~B.}, \bibinfo{author}{Bouhon, A.} \&
  \bibinfo{author}{Sigrist, M.}
\newblock \bibinfo{title}{Spontaneous surface flux pattern in chiral p-wave
  superconductors}.
\newblock \emph{\bibinfo{journal}{Physical Review B}}
  \textbf{\bibinfo{volume}{97}}, \bibinfo{pages}{064510}
  (\bibinfo{year}{2018}).
\newblock
  \urlprefix\url{https://journals.aps.org/prb/abstract/10.1103/PhysRevB.97.064510}.

\end{thebibliography}

\newpage
\clearpage
\onecolumngrid

%%% Extended Data figures %%%%

\section{Extended Data Fig. 1: Hall Data at various temperatures}

\section{Extended Data Fig. 2: Field sweeps at different in-plane angles}

\section{Extended Data Fig. 3: Field sweeps at different out of-plane angles}

%\section{}
\newpage
\clearpage
\part{Supplementary information} % Start the appendix part
%%% here starts the supplementary%%%%
\renewcommand{\thefigure}{S\arabic{figure}}
\setcounter{figure}{0}

\subsection{Low Temperature Data and Control Experiments}
Our main results (Fig. \ref{HcPhi}) is a two-fold symmetric modulation of the critical field when the magnetic field is applied in the basal plane of the \Hb{} crystal. In the paper this effect is correlated with the stripe modulations seen on the microscopic scale. %By carefully scrutinizing our experiment we have come up with
First, we show that the effect is small at low temperatures. 
Second, we rule out three possible spurious origins for the modulations in \Hc{}: current direction, sample misalignment and a wobble of the plane of rotation. 
%All those experimental factors are excluded:

\begin{figure*}[ht] %FIGS1: two cleaves + field missalignemnt
  \captionsetup[subfigure]{labelformat=empty}
  \centering
  %\subfloat[]{\includegraphics[height=6.5cm]{figures/RBsupp.pdf}\label{RBsupp}}
 \subfloat[]{\includegraphics[height=6.5cm]{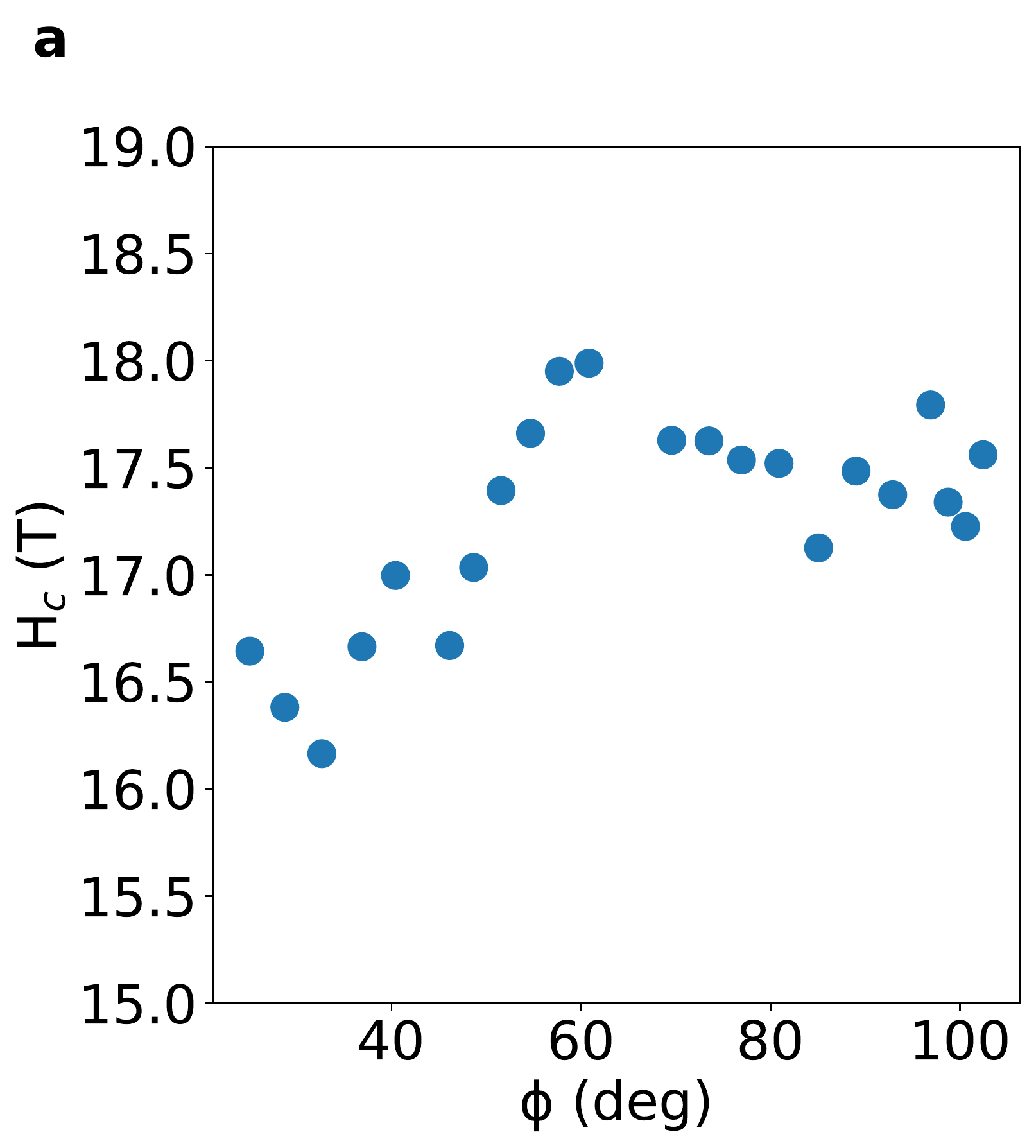}\label{HcPhiTall}}
  \subfloat[]{\includegraphics[height=6.5cm]{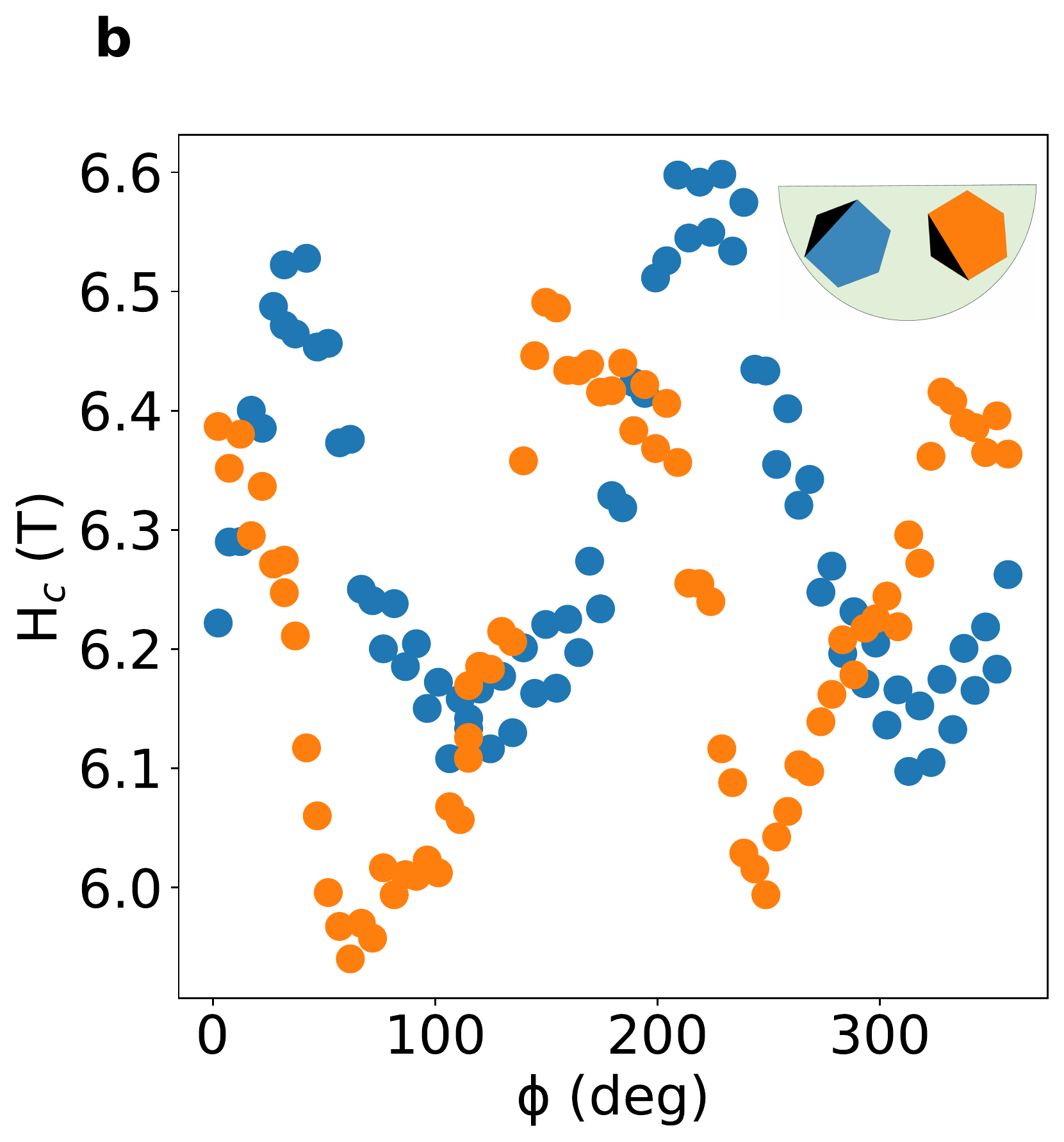}\label{HcPhiTwins}}
  \subfloat[]{\includegraphics[height=6.5cm]{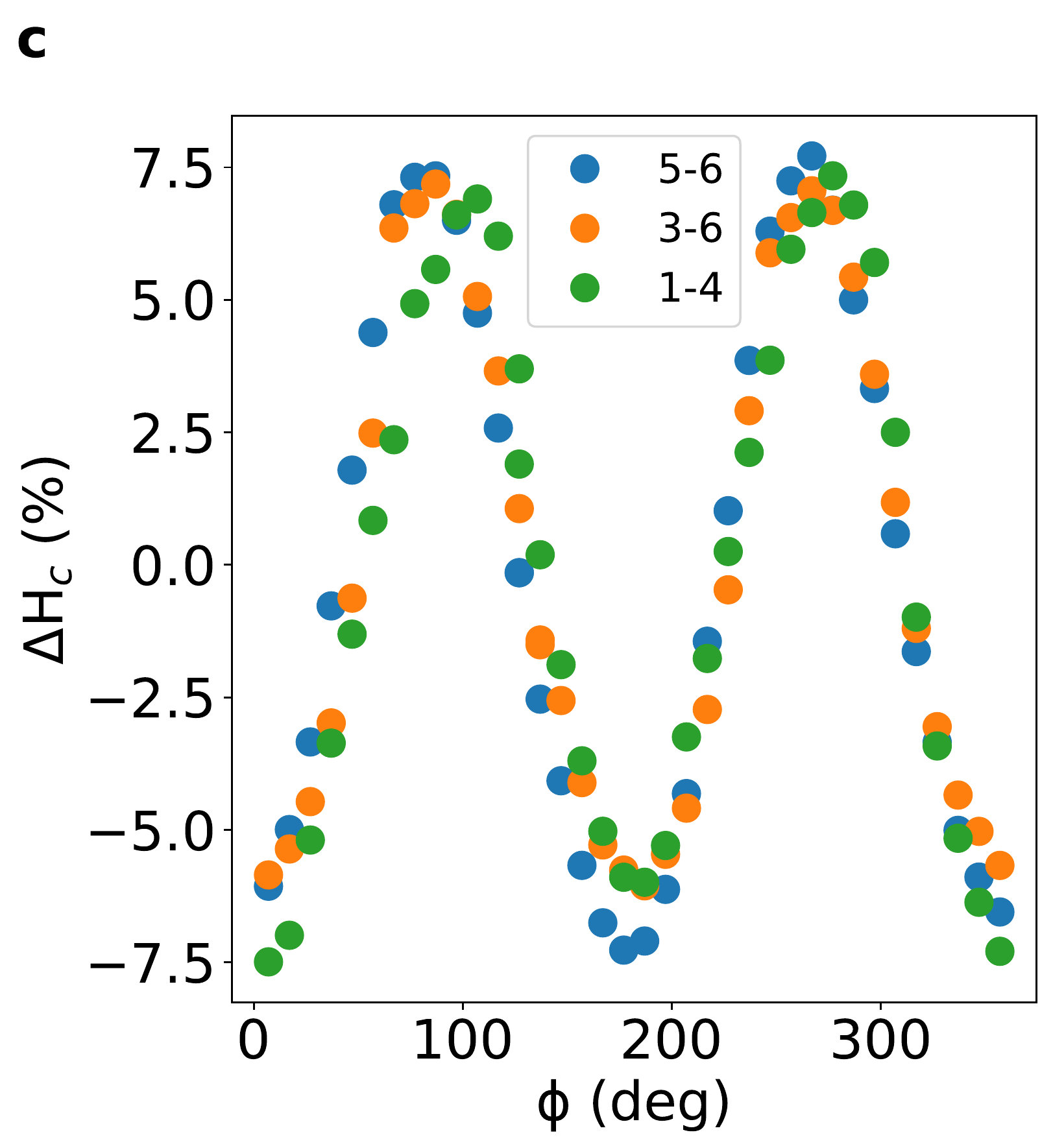}\label{HcPhiFlakediffcontacts}}

\caption{The upper critical field as a function of angle. (a) At low temperature, $T=0.4K$, a much higher field, 16-18 Tesla, is needed to suppress superconductivity. The anisotropy of \Hc{} is weaker with a variation of not more than a few percents. (b) Critical field at $T=2$~K as a function of the probe angle for two cleaves of the same crystal. The two cleaved samples were co-mounted in a relative angle of about $90^{\circ}$. Clearly the minimal critical field is shifted by a similar angle. The minimum critical field in both cases occurs when the field is applied along a crystal axes. Inset: a cartoon displaying how the two different cleaves are co-mounted on the rotating platform with a relative angle of $\simeq 90 ^\circ$. The color of the illustrated cleave corresponds to the color of the plot. (c) The critical field modulation at different contacts of the flake sample for various current directions depicted in figure \ref{FIG3} (a). Clearly, the current contacts has a negligible effect on the two-fold modulation.}
  \label{}
\end{figure*}

\textbf{Current direction.} The current direction might in general affect the critical field. To show that this is not an important effect in our data we performed an experiment in which the magnitude of the magnetic field was set to the middle of the superconducting transition. We then rotated the field in the basal plane, for four different applied current directions of the single crystal sample. We notice that the angular dependence is unchanged with the applied current direction. A finite resistance appears in directions where the critical field is the smallest independent of current direction, as can be seen at Fig.~\ref{RphiSC}. Similarly, the critical field was measured at various pairs of contacts in the flake sample. The two fold modulation is unchanged by the different contact configuration (see Fig. \ref{HcPhiFlakediffcontacts}).

\textbf{Sample misalignment.} A constant out-of-plane misalignment of an angle $\theta_{0}$ will simply reduce that critical field by a constant factor, and will not change the in-plane angular dependence. If we assume that the entire difference in the critical fields is due to an out-of-plane components, we obtain $\theta_{0} = 1.5^{\circ}$. We also note that the minimal critical field of 4.3~T is very large for $T=2.1$~K (see \cite{Ribak2020a}), meaning the magnetic field is well aligned in the basal plane. We conclude that in our experiment the maximal misalignment is small ($<1.5^{\circ}$), and it is not the source of the observed effects.

\textbf{Sample rotation wiggle.} A more subtle error could be that the out-of-plane field component changes as we rotate the sample in the basal plane. This error could arise from a possibly miss aligned rotation plane. We exclude such effect by performing control experiments. The two-fold symmetry of the critical field was reproduced for six different samples, in a total of four different rotation probes. In all of the samples the minimal critical field occurred when the critical field was pointed along a crystal axis. Thus we conclude that the effect is due to the sample properties.

Secondly, we performed a control experiment in which two cleaves of the same crystal were co-mounted in an intentional $90^{\circ}$ in-plane offset. If the variation of the critical field was only due to a possible wobble in the rotation, we would expect that the critical field will be minimal for the two samples at the same angles. However, we observed that the critical field minimum is rotated by $90^{\circ}$, meaning that the field variation is related only to the crystal and not to the (common) rotation platform (Fig.~\ref{HcPhiTwins}).( Here the critical field is defined as the field at which the resistance is 90\% of the normal state resistance).

\subsection{Proposed origin of the stripes}
We interpret the stripe pattern as a result of the layer mismatch in \Hb{}. Let us consider the geometrical aspects of the \Hb{} layers. The 1H and 1T layers have different in-plane lattice parameters with about 1.5\% mismatch~\cite{Conroy1972} and different symmetries~\cite{Narayan1976}. Without any special symmetry considerations, this mismatch will either relax isotropically or lead to a two dimensional \Moire{} pattern. However, at the charge ordered states the 1T and 1H layers form different charge density waves, leading to a non isotropic mismatch. Specifically, the CDW pattern in the 1H layer is parallel to one of the crystal axes~\cite{Scholz1982}. In this special direction the strain is maximal and the charge moves to minimize the CDW-induced mismatch.

\begin{figure}[ht] 
  \centering
   {\includegraphics[width=0.4\columnwidth]{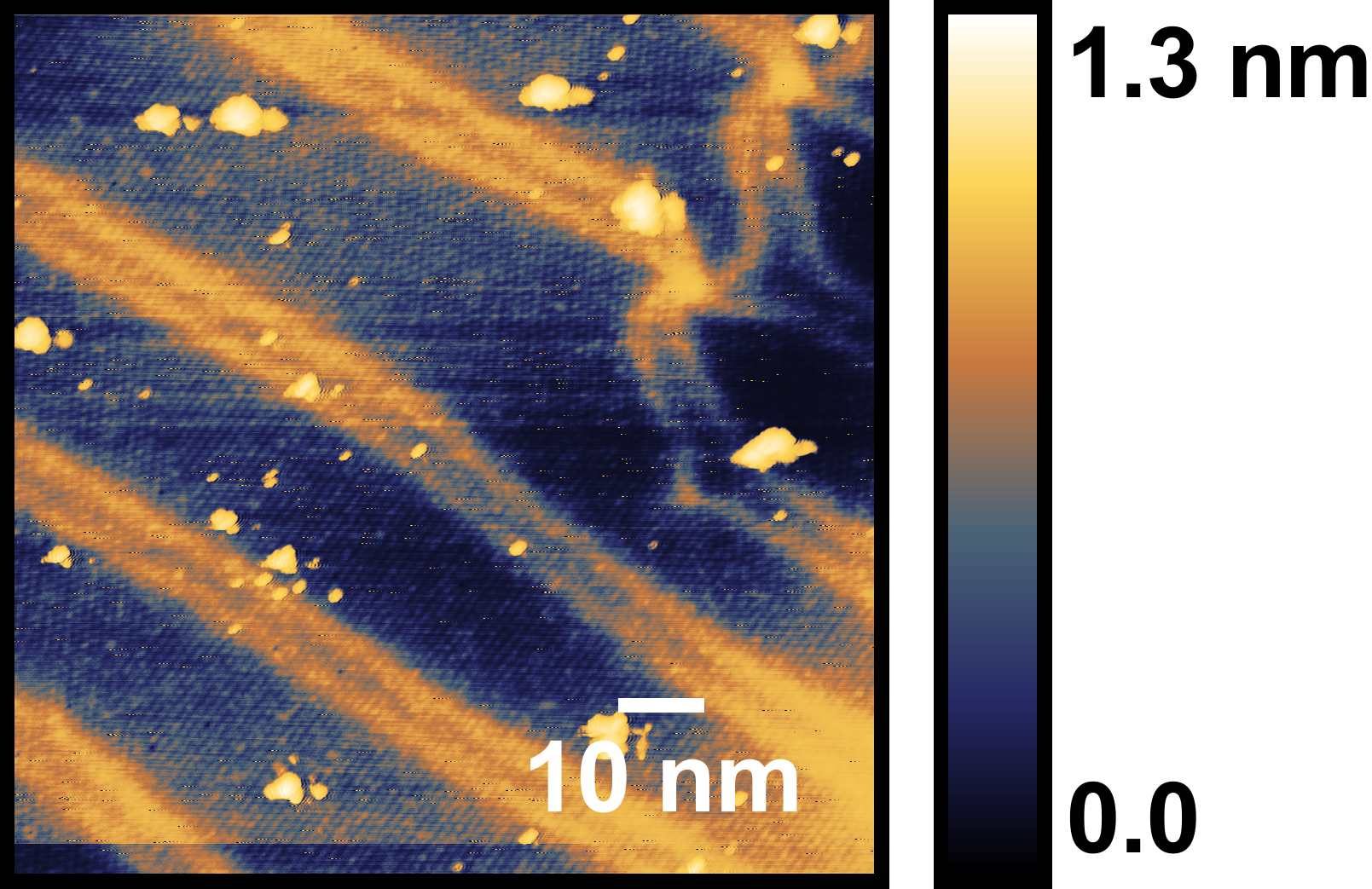}}

\caption{Stripe pattern deformation around a dislocation. Local topography measured by an STM in a region with a dislocation (top right of the image), clearly showing the bending of the stripes next to it.}
  \label{StripeDeform}
\end{figure}

This uniaxial contraction may lead to a one dimensional \Moire{}-like pattern parallel to the CDW pattern, with the expected periodicity of 19~nm due to the 1T and 1H lattice 1.5\% mismatch \cite{Ravnik2019}. The strong electron-phonon coupling in \Hb{} couples the electronic degrees of freedom to the lattice, resulting in a \Moire{}-like pattern of the local density of states, as seen in Fig.~\ref{STMstripes}. We speculate that scarce regions with no stripes are due to some local strain relaxation, as the stripes tend to deform close to a dislocation, as seen in Fig.~\ref{StripeDeform}.

\newpage
\subsection{Anisotropic Ginzburg–Landau theory with an in-plane field}
Our starting point is the  linearized GL equation
\be
\sum_{i=x,y,z}\frac{1}{2m_i}(-i \hbar \nabla_i  - 2e A_i)^2 \psi + \alpha \psi =0,
\ee
where the masses are related to the coherence length, $m_x=m_y \equiv m_{xy}=\frac{\hbar^2}{2\xi_{xy}^2 |\alpha|}$, and $m_z=\frac{\hbar^2}{2\xi_{z}^2 |\alpha|}$. Treating $H_0=\sum_{i}\frac{1}{2m_i}(-i \hbar \nabla_i  - 2e A_i)^2 $ as the  Hamiltonian of a particle in a magnetic field with eigenvalues $\{E \}$, the critical field is obtained from the minimal energy solution, $|\alpha|= {\rm{min}}\{E \}$. % while for $H>H_c$ we have $|\alpha|< {\rm{min}}E$ and there is no solution.
For the homogeneous case with a magnetic field in the $xy$ plane we have eigenstates $| k_\parallel, k_\perp , n\rangle$ ($k$ parallel or perpendicular to the magnetic field $B$ in the plane) with energy 
\be
\label{eq:he-app}
E=\hbar \omega (n+1/2)+\frac{ \hbar^2 k_\parallel^2}{2m_{xy}}+\frac{ \hbar^2 k_z^2}{2m_{z}},
\ee 
where, $\omega = \frac{2eB}{\sqrt{m_{xy} m_z}}$ so that ${\rm{min}}\{E \}=\frac{\hbar \omega}{2}$, and
\be
\label{eq:hc-app}
H_{c}= \frac{|\alpha | \sqrt{m_{xy} m_z}}{\hbar e} = \frac{\Phi_0}{\xi_{xy} \xi_{z}},
\ee 
with $\Phi_0=\hbar/2e$. For $B \parallel z$ we have $
H_{c,z}=  \frac{\Phi_0}{\xi_{xy}^2}$. Since $\xi_{xy} \gg \xi_z$ the in-plane critical field is larger than the perpendicular critical field, $\frac{H_{C,xy}}{H_{C,z}} = \frac{\xi_{xy}}{\xi_z} \gg 1$.

\subsection{Incorporation of the stripe modulation via mass modulation}
Here the anisotropy of the critical field is attributed to a modulation of  the perpendicular mass $m_z$ in a GL theory. The stripes are incorporated by adding a periodic modulation to $m_z$,
\be
m_z(x)=m_z+\delta m \cos \frac{2 \pi x}{a}. 
\ee
Here $a$ is the distance between the stripes, and we set the $x$ direction to run perpendicular to the stripes.
We treat this modulation as a small perturbation,  $\delta m \ll m_z$. The Hamiltonian used to obtain the critical field is $H=H_0+V$, where
\be
V = -\frac{\delta m}{4m_z^2}p_z^2 (e^{i\frac{2 \pi x}{a}}+e^{-i\frac{2 \pi x}{a}}) +V^{(2)}+ \mathcal{O}(\delta m^3),
\ee
Here $V^{(2)}=\frac{p_z^2}{2m_z}\left(\frac{\delta m}{m_z} \right)^2 \cos^2(2 \pi x/a)$ and $p_z$ is the momentum operator along the $z$ direction.
We next compute the correction $\delta E$ to the ground state energy. The relative correction to the critical field is 
\be
\label{eq:correction}
\frac{\delta H_c}{H_c} = -2 \frac{\delta E}{\hbar \omega}.
\ee

$\underline{\mathbf{B \parallel z}}$. For perpendicular field the screening currents flow in the $xy$ plane and are unaffected by the periodic modulation in $m_z$. Thus, $H_{c,z}$ is unaffected by $\delta m$.  Next we consider an in-plane field, $\vec{B}=B (\cos(\theta) \hat{x}+\sin(\theta) \hat{y})$.

\begin{figure*}[ht]
	\includegraphics[width=1\textwidth]{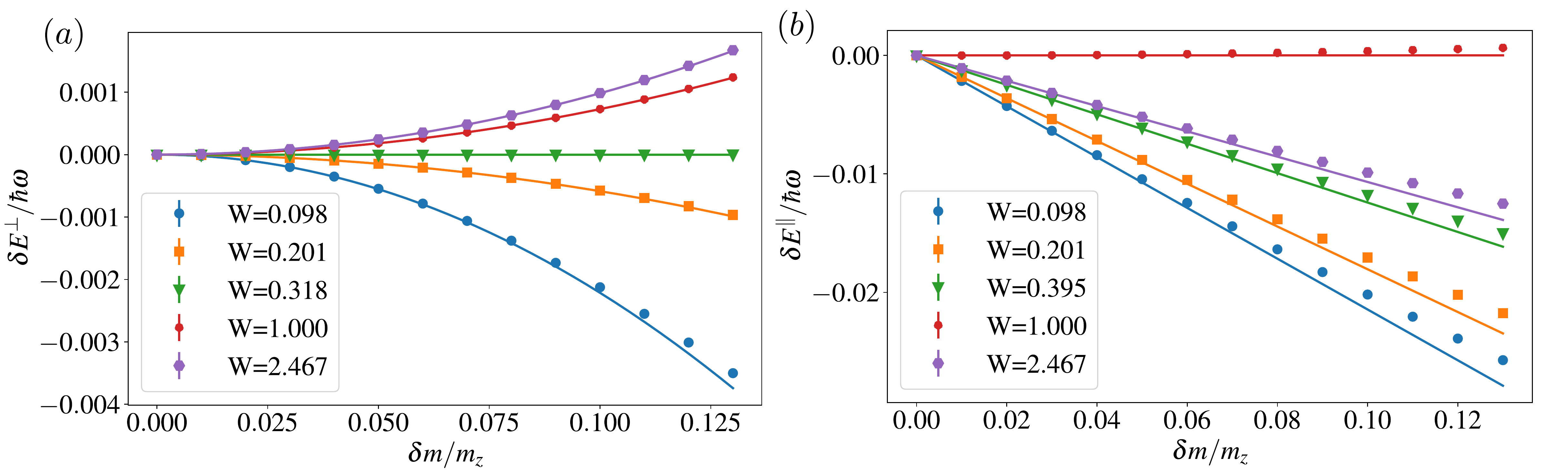}
	\caption{Correction to the lowest energy eigenvalue with increasing mass variation $\delta m/m_z$ presented for different $W$. The points are the results from the numerical calculation whereas the lines are the perturbative results. (a) $\theta=0$ that is magnetic field perpendicular to the stripe direction. The perturbative calculation predicts the correction to be second order in $\delta m/m_z$, which is confirmed by the numerical calculations. In this case, the correction is expected to change the sign at $W=\frac{1}{8}(-5+\sqrt{57})$, which is also obtained from the numerics. (b) $\theta=\pi/2$, i.e., magnetic field parallel to the stripe direction. Here the perturbation calculation predicts a first-order correction to the energy eigenvalue, which is confirmed by the numerical calculations. The first-order correction is expected to vanish for $W=1$.}\label{fig:fig1}
\end{figure*}

%Next consider an in-plane magnetic field $\vec{B}=B (\cos \theta , \sin \theta)$.
$\underline{\mathbf{B \perp {stripes}}}$ ($\theta=0$). %\subsection{$B \perp stripes$ }
%$\theta=0$, i.e. 
We have set the $x$ direction to run perpendicular to the stripes so that in the present case $B \parallel \hat{x}$, and the cyclotron motion takes place in the $yz$ plane. 
The perturbation acts nondiagonally on the momentum $|k_\parallel \rangle \to |k_\parallel \pm \frac{2 \pi }{a}\rangle$, increasing the eigenvalue of $H_0$ by $E_{BZ}=\frac{\hbar^2 }{2 m_{xy}} \left( \frac{ 2\pi}{a} \right)^2$. We define the ratio
\be
W=\frac{E_{BZ}}{\hbar \omega} = \left( \frac{\pi \ell_B}{a} \right)^2 \frac{\xi_{xy}}{\xi_z} \gg 1,
\ee
where $\ell_B=\sqrt{\frac{\hbar}{eB}}$ is the magnetic length. Then, up to second order, using $p_z = -i \sqrt{\frac{m_z \hbar \omega}{2}} (a-a^\dagger)$ we have 
$\frac{\delta  E  }{\hbar \omega}= - \frac{\delta m^2 }{32 m_z^2} \frac{1}{W} \left(1+ \frac{2}{1+2W^{-1}} \right) %\xrightarrow[W \gg 1]{~}  - \frac{3\delta m^2 }{32 m_z^2} \frac{1}{W}. 
 + \frac{\delta m^2}{8 m_z^2}$. The two terms inside the parenthesis  correspond to a virtual transition to $k_\parallel = \pm 2\pi/a$ and either $n'=0$ or $n'=2$, respectively. Here $n,n'$ denote eigenvalues of $a^\dagger a$. The last term stems from first order perturbation theory in $V^{(2)}$. This quadratic dependence on the mass modulation $\delta m$ is confirmed in Fig.~\ref{fig:fig1}(a) via a comparison to a numerical solution. From Eq.~(\ref{eq:correction}), the correction to the critical field is 
\be
\label{eq:3}
\frac{\delta H_c^\perp}{H_c} =\frac{\delta m^2 }{16 m_z^2} \frac{1}{W} \left(1+ \frac{2}{1+2W^{-1}} \right) - \frac{\delta m^2}{4 m_z^2}.
\ee

$\underline{\bf{B \parallel {stripes}}}$ ($\theta=\pi/2 $). 
 In this case 
 %For $\theta=\pi/2$ 
 the cyclotron motion lies in the $xz$ plane, and the unperturbed Hamiltonian is
% \be
% H_0=\frac{1}{2m_{xy}} \left((-i \hbar \nabla_x)^2+ \hbar^2k_y^2 \right)+\frac{1}{2m_z}(p_z+2eBx)^2.
% \ee
 \be
  H_0=\frac{(\hbar k_x-2eB z)^2+(\hbar k_y)^2}{2m_{xy}}+\frac{(-i \hbar \nabla_z)^2}{2m_z}.
 \ee
Adding the perturbation $V$, each momentum kick $k_x  \to k_x \pm 2\pi/a$ acts as a shift operator $e^{\mp i p_z \Delta z/\hbar }$ on the oscillator, where $\Delta z=\frac{\ell_B^2}{2} \frac{2 \pi}{a}$. Notice that in this case $H_0$ displays the Landau level degeneracy with respect to $k_x$. Performing degenerate first order perturbation theory, we obtain a hopping amplitude $t$ between states differing by $\Delta k_x = \pm 2\pi/a$ and having exclusively $n=0$,
\bea
t&=&\langle k_x  \pm 2\pi /a ,k_y,n=0| V| k_x ,k_y,n=0 \rangle \nonumber \\
&=& -\frac{\delta m}{4 m_z^2}  \langle 0 |e^{\mp i p_z \Delta z/\hbar } p_z^2 | 0 \rangle .
\eea
This is justified in the perturbative regime $t \ll \hbar \omega$, so that higher $n$ will appear only in second order as $t^2/(\hbar \omega)$.  
Using the Harmonic oscillator result $\langle n |e^{\alpha a^\dagger - \alpha^* a} |0 \rangle = \sqrt{\frac{1}{n!}} \alpha^{n} e^{-|\alpha|^2/2} \equiv f_n(\alpha)$, we have $\langle n |e^{\mp i p_z \Delta z/\hbar}  \frac{p_z^2}{m_z \hbar \omega/2} | 0 \rangle=- \partial_\alpha^2 f_n(\alpha = \pm \sqrt{W})$.
For $n=0$, %we obtain $\frac{t}{\hbar \omega}=\frac{\delta m}{8m_z} (W-1)e^{-\frac{W}{2}}$. The resulting 
the correction to the lowest energy is negative and given by $\frac{\delta E}{\hbar \omega} = -\frac{2|t|}{\hbar \omega}= - \frac{\delta m}{2m_z} |W-1|e^{-\frac{W}{2}}$. This linear correction is confirmed via a comparison to a numerical simulation in Fig.~\ref{fig:fig1}(b). Thus the correction to the critical field is 
\be 
\label{hcparallel}
\frac{\delta H_c^\parallel}{H_c}=\frac{\delta m}{m_z} \vert W-1 \vert e^{-\frac{W}{2}}.
\ee
Thus, this model with a varying mass along $z$ predicts an anisotropic $H_{c2}$. To find the angle at which $H_{c2}$ is maximal we compare the results for the critical field perpendicular or parallel to the stripes, Eqs.~(\ref{eq:3}) and (\ref{hcparallel}). We note that these equations carry a different dependence on $W$, determined by the typical distance between stripes. Rather than plugging in a specific value of $W$, we note that there are sizable variations of the stripe separations. We thus drop the $W$ dependencies of Eqs.~(\ref{eq:3}) and (\ref{hcparallel}) as coefficients of order unity. Thus, for small $\delta m$, the first order correction for field along the stripes is dominant, leading to a relative enhancement of the critical field, $\delta H_c^\parallel \gg \delta H_c^\perp$. 

\begin{figure*}[ht]
	\includegraphics[width=1\textwidth]{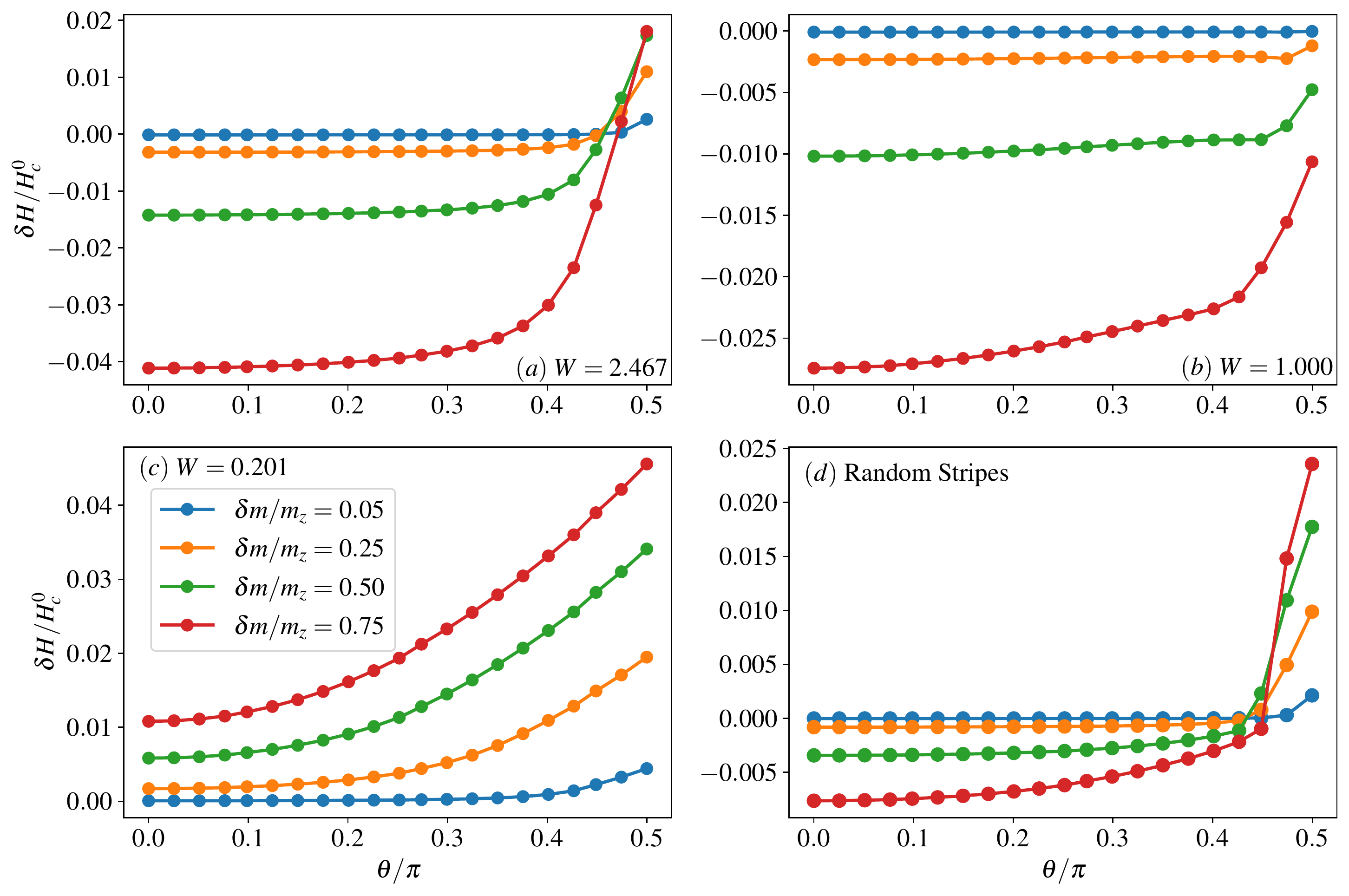}
	\caption{Critical field versus $\theta$ ($\theta=\pi/2$ corresponds to field parallel to the stripes) for various values of the mass modulation $\delta m/m_z=0.05,0.25,0.5,0.75$. (a,b,c): the calculation is done for various $W$ withing the model with periodic modulations. (d): Here we consider non-periodic mass variations as described in the text, with parameters $n=2$, $a_1=\sqrt{5}, a_{2}=\sqrt{11}$, $b_1=\sqrt{7}$, and $b_2=\sqrt{13}$.}\label{fig:ddresult}
\end{figure*}

This conclusion applies beyond the perturbative regime, and also with not strictly periodic modulations. To verify this we performed a numerical solution of the effective Schr\"{o}dinger equation away from the perturbative regime. As seen in Fig.~\ref{fig:ddresult}, for generic values of $W$ and $\delta m / m_z$ we have $\delta H_c^\parallel > \delta H_c^\perp$. We also present results from for non-periodic $m_z$ modulations in Fig.~\ref{fig:ddresult}(d). We see that the same sign of the anisotropy (maximal critical field along stripes) persists away from the periodic model as well. 

To create non-periodic stripes, we used the sum of sine and cosine functions with the irrational periodicity, which generates a profile of non-periodic stripes. The mass is now given by
$m_z(x)=m_z+ \delta m \sum_n c_n \cos(q_n x) +  d_n \sin(r_n x)$ 
where $q_n=2 \pi/a_n$ and $r_n= 2 \pi/b_n$. We select irrational $a_n$ and $b_n$, and $c_d$ and $d_n$ were chosen randomly from a uniform distribution $[0,1]$. We normalize the $c_n$ and $d_n$ as $\sum_n c_n+d_n=1$, so that the maximum variation of the stripes is given by $\delta m$. We plot the disorder averaged correction to the field for 15 independent configurations. In particularly, in Fig.~\ref{fig:ddresult}(d) we use $n=2$ and choose the following parameters: $a_1=\sqrt{5}, a_{2}=\sqrt{11}$ and $b_1=\sqrt{7}, b_2=\sqrt{13}$.

%\begin{figure*}[]
%	\includegraphics[width=1\textwidth]{figures/Fig_wtheta_random}
%	\caption{Critical field versus $\theta$ for non-periodic mass variations. Here we choose  $n=2$  with the: $a_1=\sqrt{5}, a_{2}=\sqrt{11}$ and $b_1=\sqrt{7}, b_2=\sqrt{13}$. See text for details.}\label{fig:ddresultrand}
%\end{figure*}

\subsubsection{Details for the numerical calculation}
\label{subsec:NC}
We solve the Schrodinger equation numerically for the lowest eigenstates using a grid in real space. We discretize the space in each dimension from  $[a,b]$ and use periodic boundary conditions. We choose the grid size uniformly, and hence the spatial coordinates are given by
\begin{equation}
r_i=a+\frac{b-a}{N} i
\end{equation}
where $N$ is the number of grid points in each direction, and $i$ is an integer from $1 .. N$.
The quantities that depend only on the coordinates are sampled on these grid points. The derivatives are approximated using the finite-difference approach~\cite{Varga}. We used a fourth-order approximation of the Laplacian operator. We diagonalized the Hamiltonian using the standard procedure and keep track of the lowest eigenstates. %Note that a finer mesh, i.e., higher number of grid points is expected to work as a good approximation for the continuum differential equations.
To find the continuum limit we varied $N$ from $N=101$ and up to $N=151$, and extrapolated the results to $N \rightarrow \infty$.
\subsection{Chiral order to Nematic order transition}
\label{SI:NemtictoChiral}
\begin{figure*}[ht]
	\includegraphics[width=1\textwidth]{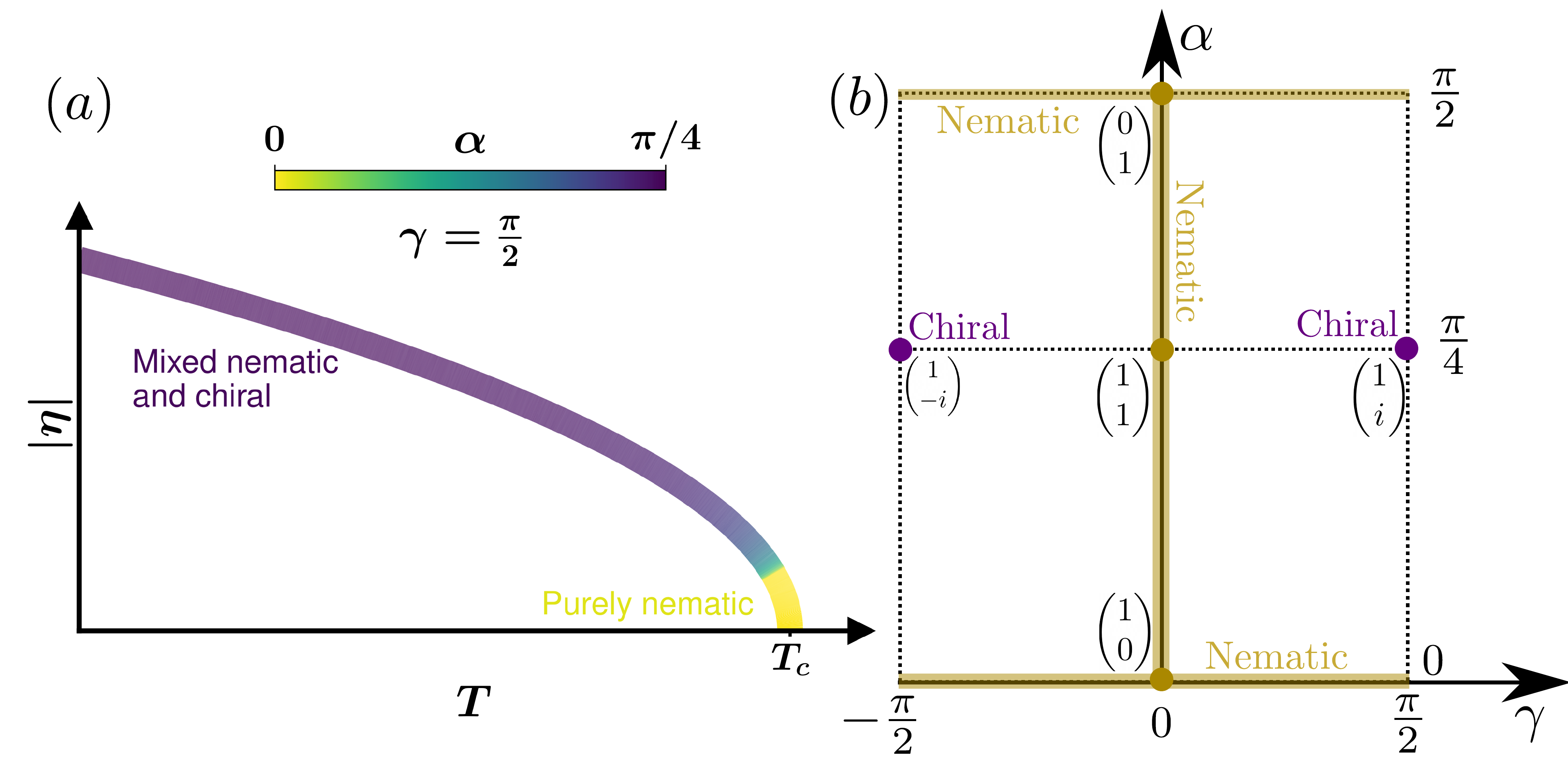}
	\caption{(a) Schematic amplitude and character (color code) of the order parameter versus $T$, followed by minimization of Eq.~(\ref{eq:homogeneous}). (b) Chiral versus nematic parametrization of the order parameter using two angles ($\alpha,\gamma$), see Eq.~(\ref{eq:aplha_gamma}). }\label{fg:chiral_nematic}
\end{figure*}

Consider the free energy of a two-component order parameter $\vec{\eta} = (\eta_x, \eta_y)^T$ ~\cite{Venderbos2016,Etter2018}:
\bea
F_\eta&=&r_\eta \vec{\eta}^* \cdot \vec{\eta} + b_\eta (\vec{\eta}^* \cdot \vec{\eta})^2+c_\eta | \vec{\eta}^* \times  \vec{\eta}|^2  %\nonumber \\ &+&
+ \kappa_1 \left( (\vec{D} \eta_x)^* \cdot \vec{D} \eta_x +(\vec{D} \eta_y)^* \cdot \vec{D} \eta_y \right)
%\nonumber \\ &+&
+ \kappa_2 \left( |\vec{D} \cdot \vec{\eta}|^2 - |\vec{D} \times \vec{\eta}|^2 \right).
\eea
Here $\vec{D}=-i \vec{\nabla}-2e \vec{A}$ and $r_\eta \propto T-T_c^{(\eta)}$ sets the onset of this two component order parameter. We ignore for simplicity the interplay of $\eta$ with a possible $s$-wave order parameter. The order parameter is parameterized as 
\be
\label{eq:aplha_gamma}
\vec{{\mathbf{\eta}}} =\eta \begin{pmatrix}
	\cos \alpha\\
	\sin \alpha ~ e^{i \gamma}
\end{pmatrix},
\ee
%such that  nematic order is achieved at $\gamma=0$, and its direction is parameterized by $\alpha$, while chiral order is described by $(1, \pm i)^T$, 
namely
\bea
{\rm{purely~nematic}}&:&\gamma=0~{\rm{or}}~\alpha=0,\pi, \nonumber \\
{\rm{purely~chiral}}&:&\alpha=\pi/4, \gamma=\pm \pi/2,
\eea
see Fig.~\ref{fg:chiral_nematic}b.

The term $c_\eta$ selects chiral order for negative values, or nematic order for positive values. We assume $c_\eta<0$~\cite{Ribak2020a}, which implies a tendency to a uniform chiral order induced at a temperature $T^* < T_c$, driven by the nonlinear coupling which is negligible at $T=T_c$. In principle, there is another, quadratic, term that promotes chiral order in the presence of a magnetic field \cite{Venderbos2016}, of the form $i B \vec{\eta} \times \vec{\eta}^*$. Such a term would promote chiral order near vortex centers but not in the bulk of the material. We neglect this term here since it is (a) typically quite small and (b) is inconsistent with the disappearance of the $H_{c2}$ anisotropy at low temperatures, see Fig.~\ref{HcPhiTall}.
%\ref{FIG3}b.
In the presence of uniaxial strain we add the symmetry allowed term~\cite{Venderbos2016}
\be
F_{\epsilon}=\lambda_{\eta} (\epsilon_{xx}-\epsilon_{yy})  (\eta_x^* \eta_x- \eta_y \eta_y^*),
\ee
which favors nematic order.

To capture qualitatively the experimental features we apply the following  procedure. We first ignore gradient terms, and for a given set of parameters $r_\eta, b_\eta, c_\eta$ and $\lambda_{\eta} (\epsilon_{xx}-\epsilon_{yy})$ we solve for the degree of nematicity versus chirality, encoded in the pair $(\alpha,\gamma)$, as well as the amplitude $\eta$ of the order parameter. This is obtained by minimizing the homogeneous free energy function 
\bea
\label{eq:homogeneous}
f_h(\eta,\alpha,\gamma) &=&r_\eta \eta^2 + b_\eta \eta^4+ c_\eta \eta^4 \sin^2 (2\alpha) \sin  (\gamma) %\nonumber \\
%- \kappa_2 B_z \eta^2 \sin(2 \alpha) \sin (\gamma) 
%&-&
- \lambda_{\eta} (\epsilon_{xx}-\epsilon_{yy}) \eta^2 \cos (2 \alpha).
\eea
While the $c_\eta <0$ term favoring chirality is quartic in $\eta$, the strain term favoring nematicity is quadratic in $\eta$. Hence, for temperature slightly below $T_c$ the system is nearly purely nematic, and gradually becomes chiral at low temperatures as depicted in the color code in Fig.~\ref{fg:chiral_nematic}a. 

To describe this explicitly, we assume for simplicity that $b_\eta \gg c_\eta$, so that the amplitude of the order parameter is minimized independently, $\eta^2=\frac{-r_\eta}{2 b_\eta}$ for $r_\eta<0$ and $\eta=0$ otherwise, as displayed in Fig.~\ref{fg:chiral_nematic}a. 
Minimizing with respect to $\gamma$ gives $\gamma=\pi/2$, so that we are bounded to the eastern vertical edge of the phase diagram in Fig.~\ref{fg:chiral_nematic}a. At $T=T_c$ ($r_\eta=0$) we have a minimum of $f_{h}$ at $\alpha=0$. Minimizing with respect to $\alpha$ gives 
\be
\frac{c_\eta (-r_\eta)}{b_\eta \lambda (\epsilon_{xx}-\epsilon_{yy})} \cos(2\alpha)=-1.
\ee
This acquires a solution only below a temperature $T^*\le T_c$ where $\frac{c_\eta (-r_\eta)}{b_\eta \lambda (\epsilon_{xx}-\epsilon_{yy})} \ge 1$. Accordingly, as shown Fig.~\ref{fg:chiral_nematic}(a), the order parameter is purely nematic for $T^*<T<T_c$, and is mixed nematic and chiral below $T^*$. Deep in the superconducting phase it is maximally chiral.

Next, given $\alpha$ and $\gamma$, we allow a spatial dependence of the form $\vec{\eta} =\eta(x,y) (\cos \alpha ,\sin \alpha e^{i \gamma})^T$. Substituting this into the gradients terms, they become $\kappa_1 F_{\kappa_1}+\kappa_2 F_{\kappa_2}$, where
\bea
F_{\kappa_1}&=& |D_x \eta|^2+|D_y \eta|^2 , \nonumber \\
F_{\kappa_2}&=&\cos(2 \alpha)  (|D_x \eta|^2-|D_y \eta|^2)
%\nonumber \\ &+&
+ \sin(2 \alpha) \cos \gamma (D_x \eta^* D_y \eta+D_y \eta^* D_x \eta).
\eea
We can see that  in the purely chiral case $F_{\kappa_2}=0$, whereas in the purely nematic case $F_{\kappa_2}$ is a simple mass anisotropy in the direction dictated by $\alpha$,  $F_{\kappa_2}=|D_{x'} \eta|^2-|D_{y'} \eta|^2$ where $\begin{pmatrix}
x' \\
y'
\end{pmatrix} =\begin{pmatrix}
\cos \alpha & \sin \alpha \\
-\sin \alpha & \cos \alpha
\end{pmatrix} \begin{pmatrix}
x \\
y
\end{pmatrix}$. Thus the  $\kappa_2$  term (denoted by $J_4$ in Ref.~\onlinecite{Venderbos2016}) is responsible for a possible anisotropy of either critical fields or of perpendicular vortices. This is dictated by a mass ratio
\be
\frac{m_{y'}}{m_{x'}} = \frac{\kappa_1 +\frac{\kappa_2}{2} \sqrt{3 + \cos (2 \gamma) + 2 \cos(4 \alpha) \sin^2 \gamma} }{\kappa_1 -\frac{\kappa_2}{2} \sqrt{3 + \cos (2 \gamma) + 2 \cos(4 \alpha) \sin^2 \gamma} }.
\ee
In the fully chiral case we recover isotropy $\frac{m_{y'}}{m_{x'}} =1$, while in the nematic case this becomes $\frac{m_{y'}}{m_{x'}} = \frac{\kappa_1 +\kappa_2}{\kappa_1 -\kappa_2}$.

We may now accommodate the experimental results within this model. When $T \ll T_c$ the order parameter is primarily chiral, $\alpha \cong \pi/4$, $\gamma=\pi/2$, so {\emph{Isotropic vortices}} will be observed. At higher temperatures, this model predicts an {\emph{anisotropic critical field}}, which is dominated by  the strain $\epsilon_{xx}-\epsilon_{yy}$. Whether the maximal critical field is parallel or perpendicular to the uniaxial strain depends on the signs of $\kappa_2$ and $\lambda$. It is expected that details of the microscopic model can yield either sign of the anisotropy of the in-plane $H_{c2}$.

Our model predicts that when the order parameter $|\eta|$ weak, nematicity is preferred. One might claim that at the center of the vortex core the vanishingly small order parameter is therefore nematic and we should expect anisotropic vortices for all temperatures. However, the STM measurement effectively probes a finite distance from the core, into the bulk, where the chiral nature of the order parameters is restored and we still expect that the shape of the vortex will represent the bulk order parameter. 

Finally, we comment on the behavior of $H_{c2}$ at  $T \ll T_c$, and the suppression of the anisotropy we found at low temperatures. To obtain the behavior of $H_{c2}$ at low $T$ it is necessary to solve the Landau level problem for the in-plane magnetic field, and then find the energy minimum by varying $\alpha$ and $\gamma$. Instead of this complicated procedure, we just show that the nematic part of the phase diagram is suppressed at low $T$. To this end, we estimate the instability of the chiral phase as a function of increasing magnetic field strength. In the chiral phase, the gap is isotropic, and at small magnetic field the gradient terms merely renormalize $r_\eta$,
\begin{equation}
    r_\eta \to r_\eta + \frac{\hbar e H}{\sqrt{m_{xy}m_z}} = -|r_\eta| \left(1 - \frac{H}{H_c}\right),
\end{equation}
see Eqs.~(\ref{eq:he-app}) and ~(\ref{eq:hc-app}). Plugging this back into the condition for the stability of the chiral phase, we find an estimate for the critical field strength
\begin{equation}
    \frac{H^*}{H_c} \sim 1 - \frac{b_\eta \lambda (\epsilon_{xx}-\epsilon_{yy})}{c_\eta |r_\eta|}.
\end{equation}
The nematic phase exists for $H^* < H < H_c$. Thus, the region of nematic superconductivity decreases sharply at low temperatures, where $r_\eta$ is large.
%To conclude, the two-component GL theory is flexible enough to  provide a consistent picture for the experimental observation of isotropic vortices and anisotropic parallel critical field by a smooth crossover from nematic order promoted by strain in the bulk, into chiral order near the magnetic vortices.
\end{document}